%% file: main.tex
\documentclass[sigconf]{acmart}
\AtBeginDocument{}

\copyrightyear{2026}
\acmYear{2026}
\setcopyright{cc}
\setcctype{by}
\acmConference[CHI '26]{Proceedings of the 2026 CHI Conference on Human Factors in Computing Systems}{April 13--17, 2026}{Barcelona, Spain}
\acmBooktitle{Proceedings of the 2026 CHI Conference on Human Factors in Computing Systems (CHI '26), April 13--17, 2026, Barcelona, Spain}
\acmDOI{10.1145/3772318.3791597}\acmISBN{979-8-4007-2278-3/2026/04}

\input{preamble}

\begin{document}

\title{Linting Style and Substance in READMEs}

\author{Hima Mynampaty}
\email{u1528521@utah.edu}
\orcid{0000}
\affiliation{\institution{University of Utah}
  \city{Salt Lake City}
  \state{Utah}
  \country{USA}
}

\author{Nathania Josephine}
\email{nathania.josephine@utah.edu}
\orcid{0000}
\affiliation{\institution{University of Utah}
  \city{Salt Lake City}
  \state{Utah}
  \country{USA}
}

\author{Katherine E. Isaacs}
\email{kisaacs@sci.utah.edu}
\orcid{0000-0002-9947-928X}
\affiliation{\institution{University of Utah}
  \city{Salt Lake City}
  \state{Utah}
  \country{USA}
}

\author{Andrew M. McNutt}
\email{andrew.mcnutt@utah.edu}
\orcid{0000-0001-8255-4258}
\affiliation{\institution{University of Utah}
  \city{Salt Lake City}
  \state{Utah}
  \country{USA}
}

\renewcommand{\shortauthors}{Mynampaty et al.}

\begin{abstract}
\input{abstract}
\end{abstract}



\maketitle
\renewcommand{\thefootnote}{}
\footnotetext{This version of this paper revises its characterization of prior work in response to post-publication feedback. Earlier versions remain available in the arXiv version history.}
\renewcommand{\thefootnote}{\arabic{footnote}}

\section{Introduction}
\readmes{} are often the first, and sometimes the only, form of documentation that users encounter when interacting with a new code base. 
Whether someone is exploring an open source library, reviewing a research prototype, reminding themselves of deployment instructions, or analyzing a dataset, a \readme{} must convey project goals, provide clear usage instructions, and set expectations for collaborations or contributors.

Despite their importance, the actual content and quality of   ~\readmes{} vary wildly in practice~\cite{liu2022readme, prana2019categorizing}.
Critical sections are often missing or disordered, language can be overly technical, ambiguous, or exclusionary, and some even lack basic navigation aids like a table of contents or internal links. 
Echoing larger trends~\cite{LYi2025LifeSciences}, accessibility elements like alternative text for images or clear link labels are frequently missing, and some files include insensitive or exclusionary phrasing. 
Issues that begin in the \readme{} can lead to poor adoption through implicit barriers to usage~\cite{mendez2018open}.

Complicating these issues is that different domains and communities have different standards for how they expect their documentation to be organized.
Patterns that suit a software library (such as PyTorch~\cite{Ansel_PyTorch_2_Faster_2024}) can be ineffective for a dataset (such as Awesome Public Datasets~\cite{awesomePublicDatasets}) or an interactive system (such as Vega-Lite~\cite{2017-vega-lite}), as different users need different levels of information about topics like data provenance, licensing, or ethical considerations. 

One solution is to lint them. 
Linters are automated evaluators of content, used to identify potential issues, enforce consistency, and improve quality and accessibility. Originally developed for programming languages, linting has since been extended beyond code to evaluate prose and technical documentation \cite{mikejordan_pacersuchow,writegood,textlint,vale}---among a variety of other domains~\cite{crisan2025linting, mcnutt2024mixing, sultanum2024data, chen2021vizlinter}. 
They allow for human-in-the-loop correction of content, highlighting issues rather than fully automating away crucial sociotechnical interactions.

However, existing documentation linters concentrate on formatting, style, grammar, and basic structural mechanics. 
For example, markdownlint~\cite{markdownlint_ruby} enforces heading hierarchy and list indentation, while Proselint~\cite{mikejordan_pacersuchow} focuses on English prose issues, such as usage and diction.
These tools are valuable for consistency and readability, yet they operate mostly at the level of syntax and style. 
As \citet{tang2023evaluating} and \citet{treude2020beyond} highlight, effective documentation is much more than merely effective English diction or semantically correct markdown usage. 
Instead, it is multidimensional, consisting of effective composition of content, structure, style, and so on.
Supporting these dimensions involves questions of accessibility, security vulnerability, or even just whether or not the code in the examples runs. 
Current linters are limited in their ability to assess these quantities beyond relatively superficial components.

In this work, we explore how we might widen the gamut of evaluable properties through a design probe called \system{}.
This system is a markdown linter based around a lightweight domain-specific language (DSL) for specifying desired \readme{} behaviors.
This DSL weaves together composable operators, including predefined ones (capturing domain-relevant functionality like filtering documents), with user-customizable operators that allow for ad hoc execution of JavaScript code and calls to LLMs. 
Together, these enable a wide spectrum of possible rules, ranging from syntactic checks (like verifying lists are formatted consistently) to semantic ones (such as ensuring that links work) to more vague checks (as in requiring that text is written unambiguously for the intended audience). 
This probe can be run from the command line as well as in an online sandbox (\systemURL{}).

We evaluate our probe in three complementary studies.
First, we conducted a (N=\numParticipants{}) user study to better understand how our design probe aligned with user expectations and desires for automated documentation feedback.
This model of evaluation is embraced warmly by users, who find it to be a useful platform on which to consider additional and new forms of rules, although they observe a learning curve in our design and nuances around how feedback should be delivered. 
Then we explored whether or not our system could be replicated through naive LLM usage. 
We find that our tool's reification of domain knowledge as lint rules provides a more consistent and detailed basis on which to evaluate \readmes{} than naive application of LLMs alone.
Finally, we demonstrated the expressivity of our approach by applying it to an alternate domain, namely culinary recipes.
We find that \system{} is pliable to this domain, and that it can easily apply extant style guides---even finding errors in common sources. 
While there are limitations to what can be achieved with this model, as higher-order notions of semantics require detailed models of the content to analyze (for instance, code semantics are well understood, but domains like recipe design are not), this approach offers a straightforward and flexible way to evaluate content.

These findings reflect our contributions.
Our work demonstrates the potential of linting the \emph{substance} of documentation and provides practical guidance for implementation that builds on familiar linting patterns, such as style. 
We find that blending well-chosen domain-specific operators, programmatic execution, and LLM calls synergistically addresses the shortcomings of each of those approaches alone.
It offers a foundation in which users can adjust rules to taste and task, as well as form new ones in a straightforward way that is amenable to computational assistance. 
While previous work has explored aspects of these strategies (\eg{} Vale~\cite{vale} includes operators, standard-readme~\cite{standardreadme} focuses on \readmes{}, lintrule~\cite{lintrule-ai-linter} uses LLMs to execute evaluations), our works extends them by bringing these threads together in a manner which supports analysis of content and not merely just stylistics. 
Through our design probe, we explore the tradeoffs encountered in linters, which we use to inform the design of future tools (\secref{sec:disco}).
This work opens the door for linters that evaluate more complex forms of documentation (such as manuals), as well as more forms of culturally mediated text-based documents (such as tutorials or game instructions).

Supplementary materials are available at \osf{} and the code is available at \github{}.

\input{figures/comparision-table-revisited.tex}

\section{Related work}

We build on prior work on documentation quality and its automatic assessment, linting, and documentation linting.

\subsection{Documentation}
Code documentation serves many purposes, including advertisement, tutorial, explanation, and reference, among other tasks~\cite{ProcidaDiataxis}.
It involves an array of varied components \cite{aghajani2020software} such as \readme{} files, tutorials~\cite{arya2023properties}, galleries \cite{yang2024considering}, blog posts, and change logs.
Assuring the quality of all of these documentation components can be both challenging \cite{terrasa2018using} and high-stakes. 
For instance, low-quality documentation can create significant barriers to understanding and participation \cite{mendez2018open}. 
Documentation is often the first point of contact a novice might have with a discipline \cite{terrasa2018using}, and so joyfully~\cite{mcnutt2025slowness} inviting them in is seen as critical for projects focused on novices.
Furthermore, with software programs and libraries becoming increasingly more complex \cite{albing2003combining}, maintaining accurate and up-to-date documentation becomes important to ensure usability, maintainability~\cite{kajko2001state}, and knowledge transfer~\cite{o2014information}. 
Echoing the adage: ``the documentation, not the code, defines what a module does''~\cite{standardreadme}.

If good documentation is then essential, then so is understanding what defines good documentation.
\citet{spinellis2010code} answers this call by assessing documentation based on the inclusion of specific structural elements, with completeness and consistency considered as important attributes. 
\citet{prana2019categorizing} collect and annotate 4,226 \readme{} files and trained a classifier to categorize their subsections into eight categories, including what, why, how, when, and who questions.
\citet{liu2022readme} conduct an empirical study of \readme{} files in GitHub repos, identifying 32 common formatting patterns which they connect with repository popularity.
We draw on these frameworks in the design of our system's capabilities, as in \autoref{tab:design-space}.

Many studies enrich these general guidelines with criteria for specific types of documentation, such as tutorials \cite{arya2023properties, petrosyan2015discovering}, APIs \cite{fucci2019using}, READMEs \cite{prana2019categorizing}, or web-based information \cite{parnin2013blogging, knight2005developing, barua2014developers}, galleries~\cite{yang2024considering}, or they may define unified qualities applicable across multiple documentation types \cite{treude2020beyond, plosch2014value}. 
For example, \citet{treude2020beyond} establish ten shared dimensions, including readability, structure, consistency, and clarity, that serve as a basis for evaluating diverse forms of software documentation.
Another criterion is community knowledge, where questions and feedback from forums, Q\&A sites, or reviews, reveal unclear areas in documentation and guide improvements to enhance understandability \cite{di2016would, barua2014developers, sarah2024improving}.
For example, \citet{sarah2024improving} collected questions from StackOverflow to identify gaps in API documentation, using these questions as metrics to both evaluate and improve documentation quality.
We build on these frameworks by combining structural completeness, quality dimensions, and community knowledge as reified by the goals guiding the design of our system \secref{sec:design-goals}.

Complementing these frameworks are tools that automatically evaluate quality.
For example, \citet{tang2023evaluating} develop a web service that can automatically assess a given documentation across a variety of metrics.
PyTextQltEval~\cite{nath2025humanizing} uses an LLM to assess the quality of documentation by evaluating metrics related to conciseness, cohesion, structuredness, traceability, and consistency.
We expand on these works by enabling users to make their own ad hoc rules rather than being bound to a predefined set of rules.

After identifying an error, a natural next step is to fix that error automatically.
Richen~\cite{shen2024richen} extracts and processes relevant information from Stack Overflow questions and automatically integrates the cleaned information into Git documentation.
\citet{zhou2018automatic} develop a framework for automatically detecting and repairing defects in API documentation.
Like these we works, we provide some automated fixers for addressing identified errors, although ours are largely mediated by LLMs.

Beyond evaluation, several works support the authoring process through design interventions or process modifications.
For instance, Sodalite~\cite{horvath2023support} supports keeping documentation up to date through a notion of automated code stories. 
\citet{horvath2022using} explore using annotations on code, an idea which \citet{misback2025codetations} extend through a hybrid edit-tracking/LLM-based method. 
While there are tools~\cite{moreno2013automatic} that provide automatic documentation generation, \citet{wu2021exploring} highlight that these approaches are no panacea as authors tend not to trust the generated results.  
Our work is complementary to these in that our linting-based strategy might be combined with those designs. 
Another orthogonal but conceptually related strategy is the use of scaffolding, in which users are provided with examples they can opportunistically~\cite{brandt09Opportunistic} modify rather than starting from a blank page. \citet{hui23lettersmith} explore such a strategy in a manner that uses structured annotations to guide users towards achieving certain goals (in the domain of letter writing).
Similarly, our work is complementary to this approach in that it could be interwoven to provide a richer construction platform.

\subsection{Linting}
One approach to identifying and addressing errors is through linting. 
A linter is composed of a collection of lint rules that each independently evaluate their subject of interest and means to execute them. 
Most commonly, linters refer to tools (such as ESLint~\cite{eslint2025}) that evaluate code stylistics, such as the maximum number of characters in a line of code or asserting that there are no unused variables. 
We refer to lint rules like \lint{no-unused-variables}.

Crucially, linters surface the knowledge that they contain in a polite~\cite{whitworth2005polite} manner: they offer contextual advice that respects the user's agency. 
\citet{mcnutt2024mixing} further conceptualized this process through the notion of a ``linting ladder'', which characterizes common linter patterns as checkable (assertions can be automatically evaluated), customizable (checks can be deactivated or adjusted), blameable (the cause of the issue can be explicitly identified), and fixable (automated corrections can be applied).
For instance, a line of code extending beyond the maximum allowed length might be fixed by indicating that the rule does not apply to a particular line, changing the allowed length, or accepting an automatically generated suggestion that splits the line.  
Linters provide structured criticism by identifying deviations from defined rules, while also supporting parameterized adjustments and fixes.

Beyond code, linters have been used in a variety of domains. 
\citet{hynes2017data} and \citet{sultanum2024data} explore linting datasets. 
\citet{mcnutt2024mixing} evaluate color palettes and identify both aesthetic and accessibility errors.  
A variety of linters \cite{hopkins2020visualint, chen2021vizlinter, mcnutt2018linting} focus on visualization, surfacing errors in various contexts and domains. 
For instance, GeoLinter~\cite{lei2023geolinter} summarizes the design guidelines for choropleth maps and creates a framework that not only finds errors in the code but also applies soft rules that choropleth maps might have.
Most closely to our own work, \citet{das2025misvisfix} explore how LLMs can be used to detect, explain, and correct misleading visualizations.
We draw on this extension to linters to radically increase the gamut of possible checks that can be evaluated. 
To that end, several systems explore the use of LLMs to execute lint rules~ \cite{naik2025metalint, Fang2025lintllm, holden2024code}. 
lintrule~\cite{lintrule-ai-linter} allows description of lint rules exclusively  through plain language.
GPTLint~\cite{fischer2024gptlint} includes positive and negative examples as means to guide LLM analysis.
Our work expands on these by interweaving LLM analysis with programmatic logic to support nuanced and specific analysis, while demonstrating the gamut of this expanded space of analysis.

These rules do not exist in a vacuum.
\citet{crisan2025linting} highlight how linters are inherently sociotechnical objects as they mediate and implement standards that are specific to a given community. 
For instance, code linters implement a specific community's coding standards, while a visualization linter implements what the (typically academic) visualization community views as an effective visualization.
This is echoed by previous linters, such as how SQLFluff~\cite{Cruickshank23SQLFluff} stresses that linters are something that a team must adopt as a group, as it implements that team's standards automatically. 
We observe a connection between documentation and linting as they are both inherently sociotechnical objects.
Linter's adaptability to a specific context is then particularly valuable for documentation, where different forms serve distinct purposes and domain knowledge often plays a critical role in assessing quality.

\subsection{Documentation Linting}
\label{sec:doc-linting}

We are not the first to observe that linters can be usefully applied to documentation. The text formulation of these documents provides a useful medium for analysis. These linters either end up evaluating textual style (\ie{} the characters that make up the file) or content style (\eg{} grammar).

Linters evaluating the textual style of markdown files are reasonably common.
Two identically named markdownlint packages (one in Ruby~\cite{markdownlint_ruby} and one in Node~\cite{markdownlint_node}) apply code linter-style checks to markdown, such as maximum line lengths and other stylistic conventions. GitHub's content-linter~\cite{github_content_linter} integrates markdownlint~\cite{markdownlint_node} with a custom rule set into its documentation infrastructure.
remark-lint~\cite{remarklint} has a similar structural overview, but tuned to the remark markdown parser.
mdformat~\cite{mdformat} is an auto-formatter for markdown that implicitly enforces this style of convention in its auto-corrections. 
readlint~\cite{readlint} applies standard code linters to all of the code examples in \readme{}-based documentation.
Closely related to our work by approach, Vale~\cite{vale} is a documentation linting platform in which rules can be developed to address specific concerns using a DSL-like structure.
The structure of our system draws on these tools (such as using an abstract syntax tree as a medium for analysis) and, as we highlight in \autoref{tab:design-space}, is able to replicate rules like \lint{require-alt-text-for-images}.

Several tools explore feedback on documentation content. alex~\cite{alex} looks for insensitive or inconsiderate writing echoing similar work from \citet{winchester2023hate}---for instance, guiding users away using language like `cripple' and towards inclusive language like `person with a limp'. 
write-good~\cite{writegood} and proselint~\cite{mikejordan_pacersuchow} enforce writing rules, such as avoiding the past tense and cliches.
\citeetals{perin2012linguistic} TextLint centers stylistic rules in natural language content, such as might appear in Strunk and White's The Elements of Style~\cite{strunk1920elements}.
The similarly named but unrelated project textlint~\cite{textlint} also operates over natural language by explicitly combining several of the aforementioned linters, such as write-good and alex. 
More abstractly, spelling and grammar checkers can be seen as linters, and so Grammarly~\cite{grammarly_website} can be seen as a representative example of such tools. 
We draw inspiration (and lint rules) from these systems.
Like these systems, we can programmatically identify specific terms and usage patterns to avoid.
However, unlike prior systems, our design flexibly provides feedback on a wider variety of content, such as \lint{introductory-paragraph-written-in-welcoming-manner}, while also being deeply customizable to taste and task. 
 
Beyond these general categories, we also find a variety of other linters that evaluate specific forms of documentation. 
For instance, Joblinter~\cite{joblint} evaluates tech job posts for issues related to sexism or cultural issues.
This linter, along with inclusivity-focused tools like alex, highlights how unconscious biases are precisely within the remit of linters: easy for some to forget about, but harmful if missed. 
Conceptually related is Textlets~\cite{han2020textlets}, which provides some lint-like affordances, such as maximum sentence length, particularly in the context of legal documents.
Commitlint~\cite{commitlint} evaluates commit messages for a number of structural properties, such as following a ``type(scope?): subject'' shape.
awesome-lint~\cite{awesome_lint} assesses the structure of a specific type of curated list, known as an `awesome' list, commonly found in GitHub \readmes{}.
Standard \readme{}~\cite{standardreadme} is an opinionated linter that defines a specification of how its authors believe \readmes{} \emph{should} be, and enforces those stylistic choices programmatically. 
Linting of specific domains can yield useful results. 
Our work is complementary to these in that we focus on linting \readmes{} in any domain, but, as we show in our expressiveness evaluation, can be adapted to specific domains.

\begin{figure*}
    \centering
\includegraphics[width=\linewidth]{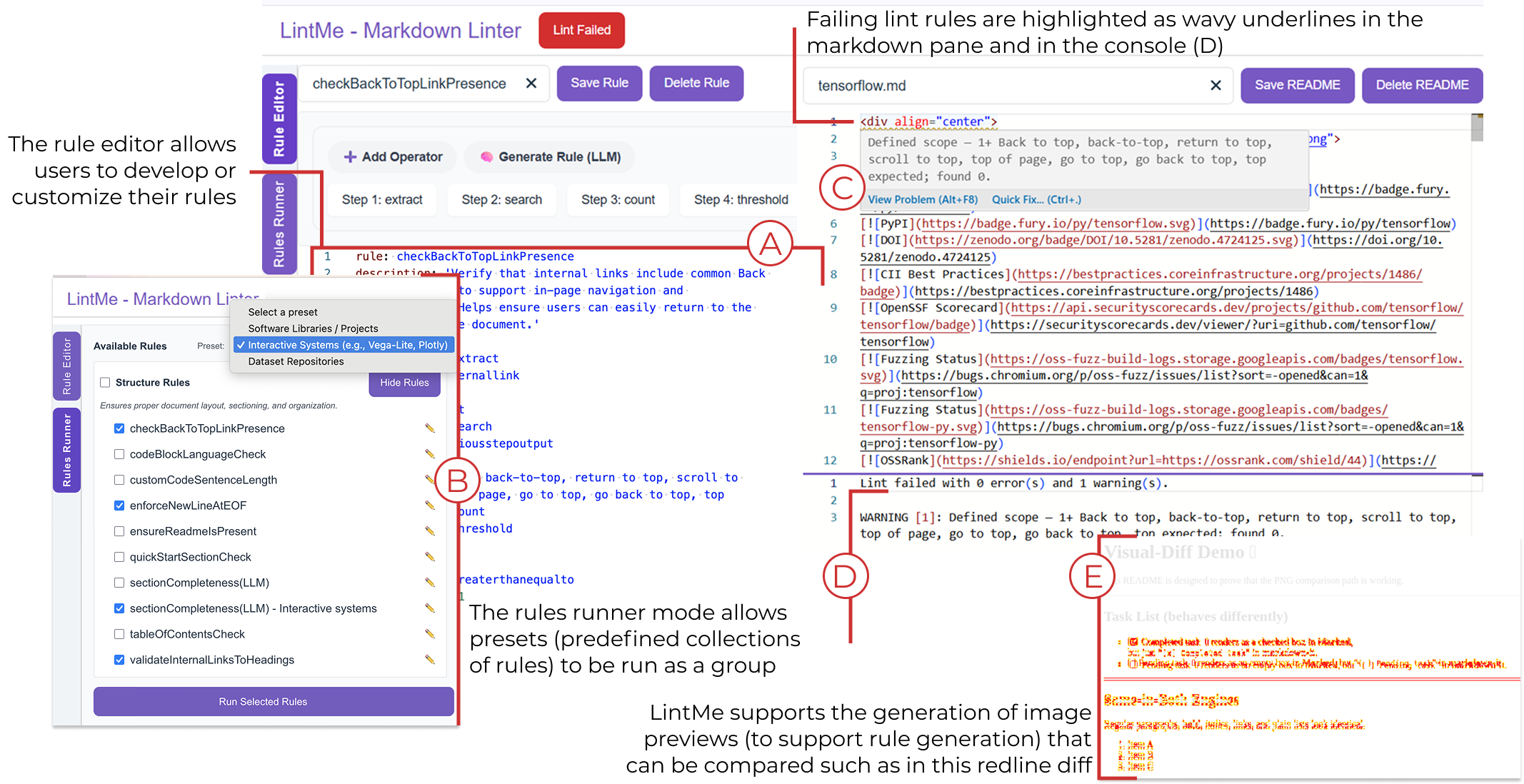}
    \caption{
    The \system{} playground supports lint rule usage as well as development and customization.
}
    \label{fig:linter-overview}
    \Description{A screenshot of the LintMe interface with annotations labeled A to E. Annotation A and B are located on the left-hand side of the interface. Annotation A shows the rule editor with the rule checkBackToTopLinkPresence and its code. Annotation B highlights the panel on the left in rule runner mode, which allows presets to be run as a group. The selected preset is “Interactive System,” with its included rules chosen.
    Annotations C–E are located on the right-hand side of the interface, which displays the README file and, at the bottom, the linting results. Annotations C and D point to the README display, showing that failing lint rules are highlighted as wavy underlines in the markdown pane.
    In addition, on the right-hand side, there is also a floating window showing the linting result, demonstrating that LintMe supports generating previews that can be compared.
    }
\end{figure*}

\section{System Design}

We now describe the design and implementation of \system{}. 
We begin by outlining the design goals that shaped our system in \secref{sec:design-goals}, then describe how those goals manifest themselves as our probe in \secref{sec:system-design}. 
To situate these richer descriptions, we first sketch the usage model for the system. 

\paragraph{Example Usage}
A user interested in checking their \readme{}, goes to our online playground at \asLink{https://lintme.netlify.app/}{lintme.netlify.app}. 
They copy their \readme{} into the text pane (\figref{fig:linter-overview}C) and, in the `Rules Runner' mode (\figref{fig:linter-overview}B), they select an appropriate collection of rules, called a preset. 
They click `Run Selected Rules,` which then evaluates all of the relevant rules selected. 
A number of the rules fail, which is summarized both in the rule selection area, in the console beneath the text entry area (\figref{fig:linter-overview}D), and on top of the markdown itself. 
Within the text entry area, errors are highlighted as red or yellow wavy underlines (\ala{} spell check), which they can hover over to see the identified issue. 
They can then iteratively adjust and correct the identified issues, regularly re-running the rules as needed. 
One of the failing rules, \lint{ensure-neutral-tone} catches their eye, and so they click customize. 
They find that the rule does not capture some of the nuances of what they would expect from a neutral tone, and so they modify the rule by altering its YAML code (\figref{fig:linter-overview}A). 
Satisfied with their changes, they run the rule again and return to modifying their \readme{}.

\subsection{Design Goals}
\label{sec:design-goals}

To guide the design of \system{}, we  identified three design goals related to our problem framing: \emph{rules should be (easily) authorable} (\gAuthorable{}), \emph{rule application should not be domineering} (\gAgency{}), and finally, \emph{they should be usable by a variety of software communities}. 
These design goals were developed as polite~\cite{whitworth2005polite} complements to the steps of the linting ladder~\cite{mcnutt2024mixing}, which asserts that a linter should be \emph{checkable}, \emph{blamable}, \emph{adjustable}, and have automated \emph{fixes}.

\subsubsection{\gAuthorable{} --- User authorable}
There is a large landscape of possible rules for a \readme{} linter, and we expect there are many others yet unidentified, including those limited to specific software communities. 
Users should be able to add their own rules that are adapted to their own specific contexts and usages.
Creation of these rules should be sufficiently straightforward that expertise in a separate set of systems is not necessary (motivating our use of LLMs rather than a deep evaluation language), and understanding the way the rules operate is similarly straightforward.
This echoes prior work on end-user manipulable feedback mechanisms~\cite{han2020textlets}, and is specifically aligned with the value of reifying~\cite{beaudouin2021generative} community standards as lints---in tandem these yield our DSL-centered approach.

\subsubsection{\gComm{} - Community-Tunable Rules.}
Instead of prescribing a single canonical style of rule, rules should  be adaptable to local practices, goals, and levels of strictness---echoing the inherently sociotechnical nature of lint rules~\cite{crisan2025linting, Cruickshank23SQLFluff}. 
For instance, interactive systems might center demos and visual displays, dataset repositories may value citation form or metadata, and libraries may emphasize runnable snippets and quick starts. 
Similarly, while a closed-source organization might use shorthand or obscure project names, such practices might not be appropriate in large open-source projects. 
Teams and communities should be able to build on shared rule sets, customize them to their needs, and revisit those choices over time as practices change. 
This captures the necessity of customization and the necessity for breadth in the gamut of possible lint rules.

\subsubsection{\gAgency{} --- Usability and Agency.}
Finally, the system should be smooth but not frictionless~\cite{mcnutt2025slowness}, so that improving documentation quality is easy and decisions remain intentional.  
For instance, a completely automated \readme{} fixer would likely push users into accepting whatever was generated due to automation biases~\cite{heer2019agency} (broadly, the human tendency to overvalue automatically generated results), thereby effectively reducing intentionality. 
While interactions should reduce the cognitive overhead of evaluating documentation quality, agency must be encouraged and retained.
Lastly, these rules should, as much possible, not impose limitations on what can or can not be evaluated. A system that enables some checks while disabling others does not respect a user's agency to evaluate what is important to them in a \readme{}---a commitment which yields \system{}'s ability to evaluate substance as well as style.

\subsection{\system{} Design}
\label{sec:system-design}

We describe the design of \system{} and note how it accomplished our design goals. 
First, we describe the core abstraction in \system{}, the \emph{operators}, which are composable building blocks used to define linting rules.
Then, we present \system{}'s interfaces for authoring rules from operators, including support for debugging and testing. 
Finally, we describe the process of running lint rules in \system{} and how the results are presented.

\begin{figure}[t]
    \centering
    \includegraphics[width=\linewidth]{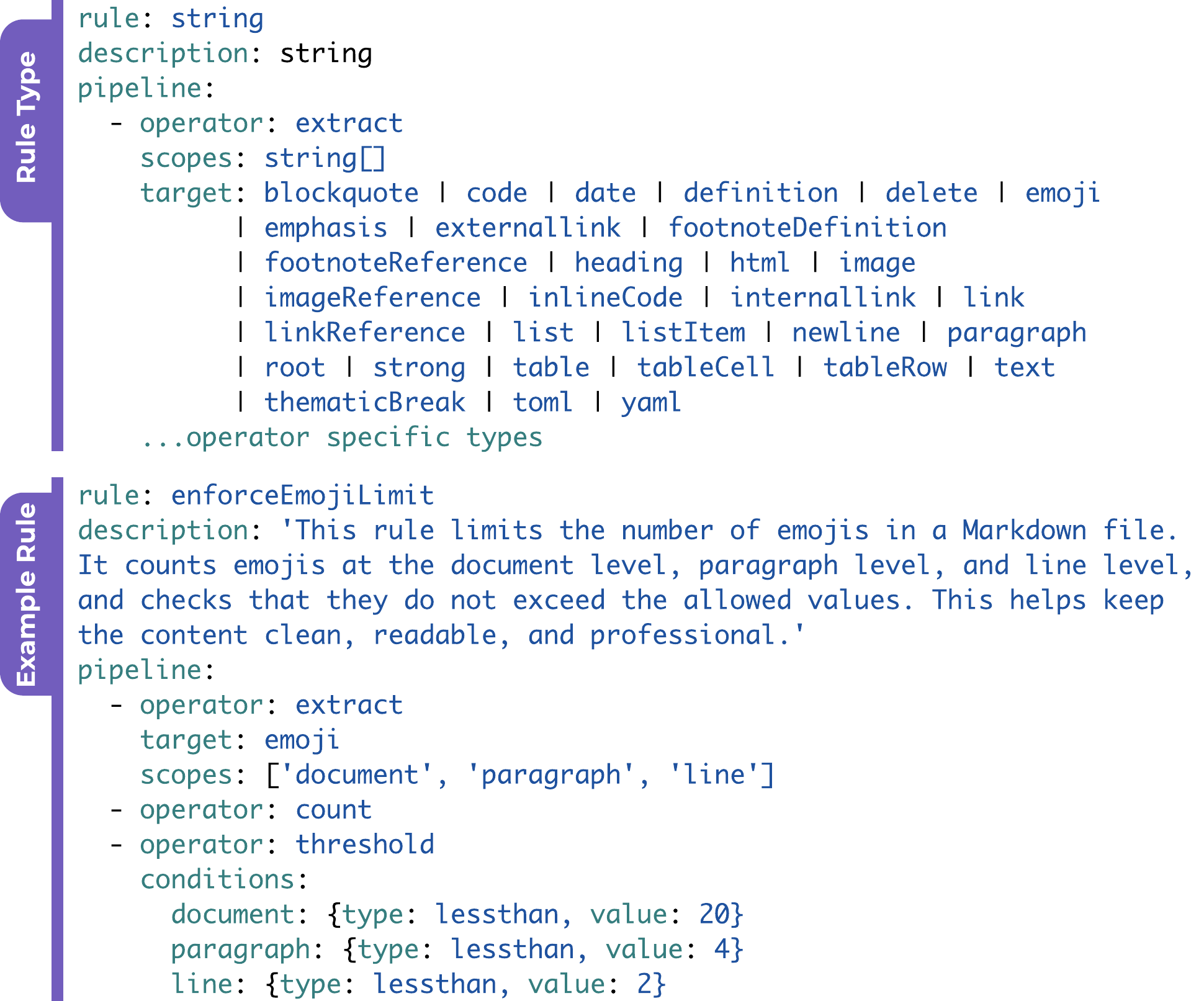}
\caption{Each \system{} lint is a configurable pipeline of composable operators, each operator possessing a unique typing. 
    For instance, \operator{extract} targets any mdast AST node  (\eg{} delete being \st{styled text}) scoped to one of several regions, and returns a list of results.
}
    \label{fig:rule-type}
    \Description{A code example showing how linting rules are defined within LintMe. The top section is labeled “Rule Type” and serves as a placeholder for rules, with the rule name and description defined as strings. The pipeline consists of an extract operator, scopes strings, and multiple possible targets for example blockquote, code, date, definition, delete, emoji, emphasis, etc.

    The second section, labeled “Example Rule”, provide an example of a rule named enforceEmojiLimit. It’s description stated that “This rule limits the number of emojis in a markdown file. It counts emojis at the document level, paragraph level, and line level, and checks that they do not exceed the allowed values. This helps keep the content clean, readable, and professional.” 
    
    The pipeline includes three operators, the first one, extract( with target set to emoji and scopes a set consisting document, paragraph, and line. The second operator is count. The third operator,  threshold (with conditions specifying that within the document the threshold is lessthan 20, paragraph less than 4, and within line less than 2).
    }
\end{figure}

\subsubsection{Rules and Operators} Lint rules in \system{} are short pipelines of \emph{operators}. 
Operators are named building blocks designed to perform specific tasks on the markdown. 
They are chained into pipelines to perform the checks defined by rules. \figref{fig:rule-type} shows the schema of the rule type and an example rule.

The example rule in \figref{fig:rule-type}, \lint{enforce-emoji-limit}, defines an upper bound on how many emojis can be used in a \readme{}. 
It is composed of a pipeline of three operators: \operator{extract}, \operator{count}, and \operator{threshold}. 
The \operator{extract} operator generates lists of a given target from the markdown, which in this case is emojis. 
It can be applied across scopes, including the document as a whole, paragraphs, or lines of text. The \operator{count} operator then produces counts for the lists generated by \operator{extract}.
Finally, the \operator{threshold} evaluates the values generated by \operator{count} against each scope's limits.

The short pipelines that define each rule constitute the backbone of \system{}'s domain-specific language (DSL), with each rule consisting of the pipeline and relevant metadata. 
\system{} provides 21 operators, a sample of which is listed in \autoref{tab:operators}. 
The full list of operators is available in our \asLink{https://lintme.netlify.app/readme.md}{system documentation}.

Many operators are parameterizable, increasing their reusability to a variety of different contexts (\gComm{}).
For instance, specifying different values for the parameter ``target'' of the \operator{extract} operator allows it to support both \figref{fig:rule-type}'s emoji threshold, as well as a rule setting the maximum number of bullet points in a list or the maximum number of headings. 
This level of composition also allows for straightforward adjustment or tuning (\gComm{}), as users only need to adjust parameters of interest to fit rules to their specific context, \eg{} setting higher threshold values for allowed emojis in our \lint{enforce-emoji-limit} rule.

Our goal in designing these operators was that they be abstract enough such that they can be composed into original rules, but not so abstract that users would be essentially using another general-purpose programming language (\gAuthorable{}).
A key design decision is then how to appropriately balance the level of abstraction in the granularity of our building blocks (operators). 
At one end of this spectrum, operators might contain entire lint rules.
For instance, a \lint{table-of-contents} rule might simply check if there is a table of contents and return a boolean---however, this would impede users' ability to author their own rules (\gAuthorable{}). 
At the other end of the spectrum, operators could provide no structure, meaning that a rule creator might need to fashion the check entirely from scratch---which would limit users' ability to do the things they want~\gComm{}.

We arrived at this collection of operators through a process of iterative refinement. 
We initially implemented a collection of lint rules from previous documentation linters (\secref{sec:doc-linting}) as single-purpose functions, again like a \lint{table-of-contents}.
We then identified common functionality between them and re-implemented the rules using the operators. We repeated this process several times until we hit conceptual saturation.  
Operators that completed multiple forms of functionality were broken apart. For instance, we considered a count item operator (\eg{} \operator{countItem("just")}), which we separated into \operator{count} and \operator{extract}. 
Other granularities and combinations of operators are possible, but this one seemed to be well matched with our domain \gAuthorable{}.

While our operators cover a range of existing rulesets, we recognize that there are desired rules beyond what we have anticipated that may be difficult or impossible to express with the provided set of operators. To support these situations,  we include several conceptual ``escape hatch'' operators that enable the lint creator to create ad hoc functionality driven by their task, further supporting \gAuthorable{}. 
The first is the prompt-based \operator{EvaluateUsingLLM}, which arbitrarily executes prompts with context provided by the chained operator. This operator allows lint creators to write rules for stylistic concerns that would be difficult to implement by more common means, \eg{} given a sentence for stylistic errors (\eg{} \operator{EvaluateUsingLLM}(``This sentence should be written in a kind manner'').
Another escape hatch operator is \operator{customCode}, which enables execution of arbitrary JavaScript (JS) code. 
This operator supports bespoke logic, such as elaborate conditionals, as well as the reuse of established functionality from JS's sprawling library ecosystem.
For example, a lint might require that \readmes{} are \lint{readable-at-a-5th-grade-level}. 
Rather than re-implementing readability metrics, such as the Flesch-Kincaid~\cite{kincaid1975derivation} readability score, the \operator{customCode} operators allow the author to simply use library code implemented previously~\cite{fleschKincaidWords}.
Lastly, our \operator{execute} operator runs identified commands on the command line, enabling evaluation of installation instructions and other pieces of code.

This operator-based design draws on a range of DSLs, including relational algebra and data-serialization format-based DSLs~\cite{mcnutt2022no} like varv~\cite{borowski2022varv} or Vega-Lite~\cite{satyanarayan2016vega}.
Among them,  Vale's~\cite{vale} small configuration language for describing lint rules is most closely related to our approach.
We extend this notion by including a wider array of operators, supporting a wider gamut of possible checks.
This design also bears a conceptual resemblance to \citeetals{laurent2025oracular} notion of oracular programming, which interweaves LLMs deeply into the notion of coding itself. 
Whereas their work is more general, ours is specific to the domain of lint-like evaluation of markdown files. 
We draw on these pliable designs to support the malleability of our rules (\gAuthorable{}) as well as the scope of things that can be evaluated (\gAgency{})

\input{figures/operators-table.tex}

\subsubsection{Operator Implementation} Here, we separate a variety of implementation decisions that are orthogonal to the design decisions guiding the construction of our DSL. 

Each operator implements its own evaluation independently. For instance, if two \operator{extract} operators are in a pipeline, they will both conduct their own abstract syntax tree (AST) parse. 
Like other markdown linters, we make use of the mdast~\cite{mdast} markdown parser to generate an AST. 
This decision was made for implementation simplicity, but could be optimized in the future.
In particular, adding a global cache that shares process information (such as markdown parses or LLM calls), as well as complete operator evaluations (such that common functionality might be reused across rules), is a promising approach.  
This modular strategy is centered on evaluating markdown files, but it also supports the consideration of metadata.
For instance, the \operator{FetchFromGithub} enables access to the git history, enabling potential rules about recency, such as \lint{\readme{}-must-be-updated-in-the-last-year}, or blame, such as \lint{\readme{}-should-exclude-all-contributions-from-a-specific-person}. 
The latter may be desired in situations where a contributor is known to act maliciously or otherwise without sign-off from the organization.

Operators are implemented as JS functions that operate over a shared context object that is sustained for a single rule execution. 
Between operators, there are a limited number of types that are exchanged. 
These include \emph{Markdown documents}, which are the raw README string,
 as well as their corresponding ASTs (as generated by mdast).
There are \emph{Scoped extractions} (such as resulting from \operator{Extract}), which are collections of matches grouped by scope (\eg{} \texttt{document}, \texttt{paragraph}, or \texttt{line}).
For instance, running just the \operator{extract} operator in \figref{fig:rule-type} yields extractions for the document (containing all AST nodes of emoji use in the document), paragraph (AST nodes of emoji usage per paragraph), and line (AST nodes of emoji use grouped by line).
Complementing these are metadata, such as 
\emph{Metric summaries}, which are numeric aggregates like counts or lengths that are derived from extractions per scope, and \emph{Diagnostics} are line-numbered messages with severities.

\system{} uses a simplistic duck-typing system, where operators attempt to coalesce whatever is passed to them into an appropriate form. 
For instance, operators such as \operator{extract}, \operator{regexMatch}, and \operator{search} then attempt to consume whatever is given to them as an \textit{AST} or coalesce it into an AST, yielding scoped extractions.
Aggregators like \operator{count} or \operator{length} implicitly collapse these collections into numeric \textit{Counts}, which enforcement operators, such as \operator{threshold}, read to emit \textit{Diagnostics}.
In light of these type coercions, operator order is non-commutative, meaning that changes to the order can yield significantly varied effects. This is broadly typical for pipeline-style structures, such as in P6~\cite{li2020p6}.
An alternative to this design might have used declarative constraints (echoing prior work's predicate-based linting~\cite{mcnutt2024mixing}), however, this pipeline model addressed our properties of interest in a simple to implement and understand way.

\begin{figure}
    \centering
    \includegraphics[width=\linewidth]{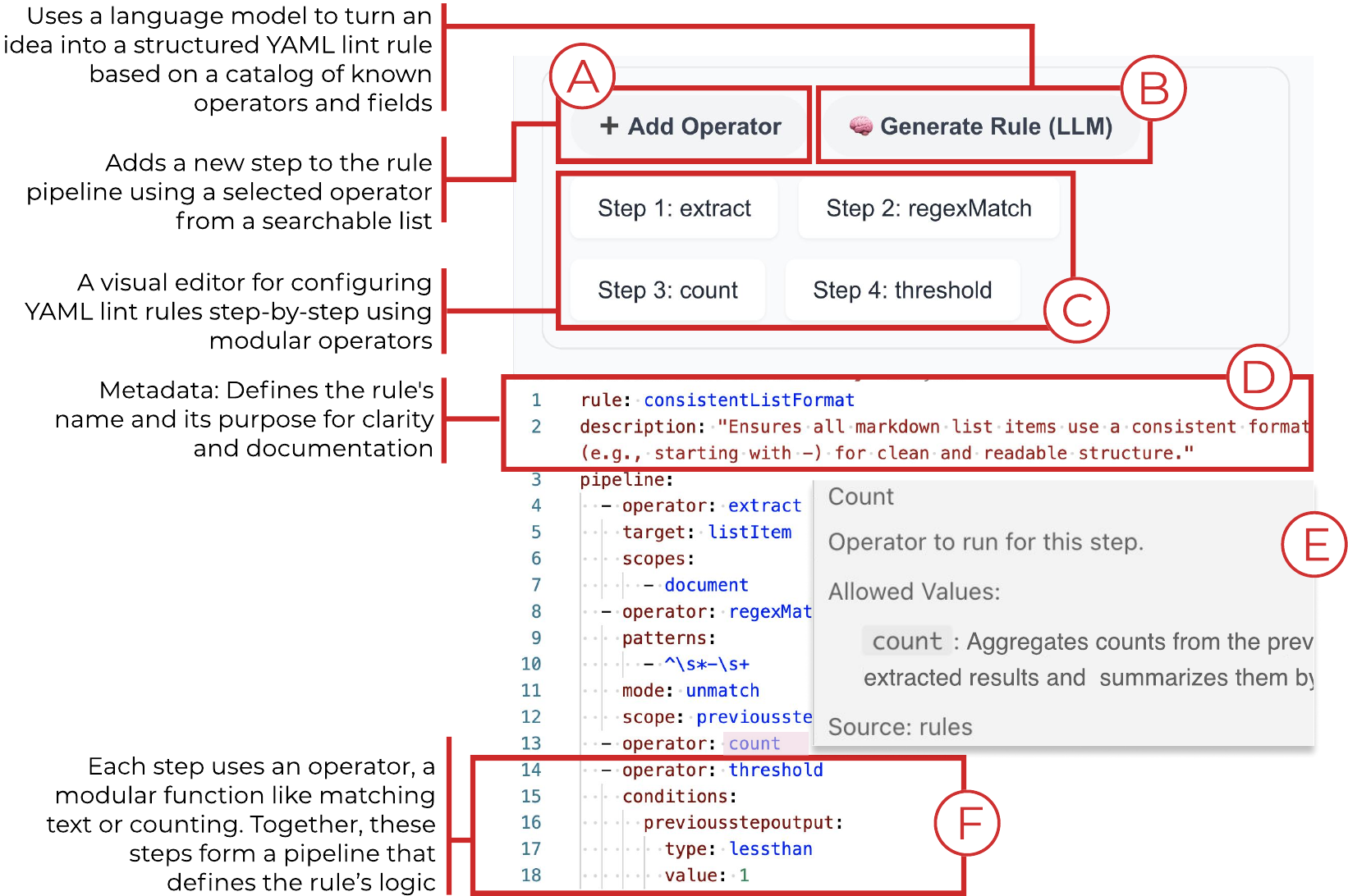}
\caption{
    Our playground includes a variety of different types of support for authoring or customizing lints, including inline documentation (E) and automatic rule generation (B). 
}
    \label{fig:labeled-rule}
    \Description{A screenshot of a part of the LintMe interface, annotated from top to bottom from A–F with a description for each. Annotation A highlights the add operator button with the description “Adds a new step to the rule pipeline using a selected operator from a searchable list.” Annotation B highlights the generate rule (LLM) button with the description “Uses a language model to turn an idea into a structured YAML lint rule based on a catalog of known operators and fields.”
    Annotation C highlights the four step buttons located in the visual editor with the description “A visual editor for configuring YAML lint rules step-by-step using modular operators.” Annotations D–F are located on an example rule description for the rule consistentListFormat. Annotation D highlights the rule name and description of the rule, with the description “Metadata: Defines the rule’s name and its purpose for clarity and documentation.” Annotation E highlights a tooltip that shows additional information on an operator. Annotation F highlights an example of the operator threshold, which is the last operator of the rule. The description reads “Each step uses an operator, a modular function like matching text or counting. Together, these steps form a pipeline that defines the rule’s logic.”
    }
\end{figure}

\subsubsection{Support for Creating Lint Rules}
With our operators and rule DSL in hand, we can now assemble them in lint rules. 
\system{} can be run as a command-line tool (in the manner of many linters) or in a web-based sandbox.
This sandbox, available at \systemURL{}, supports developing and executing lint rules. 
The web interface of \system{} has two main views (\figref{fig:linter-overview}), the editor and rule collection on the left, and the linting results on the right. 
In the design of the playground, we strive to ensure that users retain their agency (\gAgency{}) over the documentation authoring process by making the tool deeply adjustable.

A new lint rule is written by inputting YAML directly in the editor (\figref{fig:labeled-rule}).
Users may start by inputting an existing rule and modifying it, addressing the blank page problem~\cite{mcnutt2021integrated}.
In addition to manual input, users can also search the list of available operators using the \emph{Add Operator} popover (\figref{fig:labeled-rule}A) or use the \emph{Generate Rule (LLM)} functionality (\figref{fig:labeled-rule}B) to help generate a rule from natural language text. 
LLM rule generation involves a straightforward call to LLM, with a prompt that includes the full extent of \system{}'s documentation.
Users can select from one of several LLMs, including llama-3.3-70b-versatile, llama-3.3-70b-versatile, and openai/gpt-oss-120b (which are provided via the groq~\cite{groq} LLM cloud portal). 
Models are set to 0 temperature where available to reduce non-determinism. 
This selection of models and their versions is mediated by our LLM provider in our probe.
In future work we intend to expose model selection for CLI users, so that users can select whatever models (archival or not) that they wish. 
See supplement for details. 
Once satisfied with a rule, the author can save it, which adds it to the collection of rules available to run.
This design enables users to modify or create rules adapted to their specific domain or area (\gComm{})---although the automation risks automation bias (\gAgency{}).

It can be challenging to author a program of any kind, \system{} rules included, and so we provide additional features for authoring and debugging lint rules (\gAuthorable{}). 
The first of these is inline documentation via a JSON Schema. This schema automatically provides type hints on hover, as in \figref{fig:labeled-rule}E, and provides a means to automatically check the rules for syntactic correctness. 
This tightly integrated feedback is designed so that lint authors can avoid trips to the documentation to check the correctness of their designs.

Another tedious task in writing programs is mentally evaluating the instructions of partially-complete and not-yet-executable code to check its logic and correctness. 
\system{} allows incomplete programs to be executed, thereby assisting authors as they reason about their code. 
When incomplete programs are being executed (identified by the programming not ending in a boolean, like \figref{fig:labeled-rule}F), \system{} generates a warning that the program does not yield a judgment.
In those cases, it either returns a stringified version of the output of the last step in the rule to the console (\figref{fig:linter-overview}D) or generates an image and provides a link to it (as in \figref{fig:linter-overview}E) if relevant.

\subsubsection{Running (a collection of) Lint Rules}
Next, we describe the process and experience of running the lint rules.
A given lint rule is executed in several steps. 
First, the types of the operators involved are checked---for instance, a \operator{count} with nothing preceding it will halt as unevaluatable.
Once checked, each of the operators are executed in turn, roughly following a stack-based evaluation style, such that one result is fed into the next as appropriate. 
Then the result of the lint is summarized in a way that it can be integrated into the playground.
For a failing rule, this summary includes a message explaining what went wrong and what caused the error.

A critical part of linters is being able to form groups of lint rules that as a whole implement the standards of a particular community (\gComm{}). 
\system{} has a notion of presets (\figref{fig:linter-overview}B) that include a selection of different lint rules that are relevant to that domain.
In our design probe, we include three presets: software library, interactive system, and dataset repository. 
For instance, in addition to basic readability lint rules (\eg{} \lint{table-of-contents}), the dataset preset includes rules specific to its contexts (\eg{} \lint{citation-bibTeX-present}). 
These preset rule sets are currently authored through code (and so authoring new ones requires modifying the library itself), but in the future, we intend to materialize them as shareable JSON bundles. 
These bundles might be usefully mediated through a rule repository that allows users to publish rules for reuse (particularly ones that pass through a validation process). A key challenge in such a repository is the distribution of rule updates. This could involve npm-style tagged versioning (supporting version locking), or more dynamic mechanisms, \ala{} ActivityStreams or ActivityPub, which might also afford more social maintenance.
We leave exploration of these designs to future work, as the simple bundles used in our probe are sufficient to demonstrate the concept.

\subsubsection{System tour}
Finally, we give a tour of \system{} through the lens of \citeetals{mcnutt2024mixing} linting ladder.

The most primitive part of a linter is its execution, or whether or not it is \emph{checkable}. 
As in \figref{fig:linter-overview}B, the linter takes in lint rules and runs each of them for a given input \readme{}.

\begin{figure}[t]
    \centering
    \includegraphics[width=\linewidth]{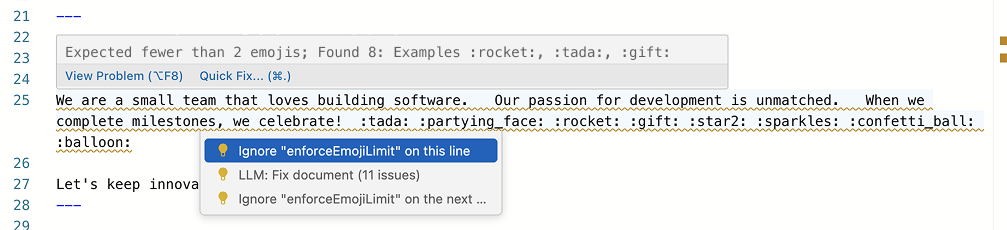}
    \caption{\system{} can identify (\ie{} \emph{blame}~\cite{mcnutt2024mixing}) the specific line or text chunk that caused a specific error. If a specific item can not be identified, then the warning is placed at the top of the file. Like lint errors in code, this error can be ignored, or an attempt can be made to automatically fix it.
}
    \label{fig:localized-error}
    \Description{A screen shot of the markdown editor from the LintMe playground. It shows a highlighted error for paragraph with too many emoji. The line is marked by a wavy-underline akin to a spell checker. Above the line is the effect of hovering on that line which details the error. Below it is the effect of clicking the line which shows a series of possible actions to take in response to the error, including ignoring it for this line and automatically fixing it.  }
\end{figure}

Next is \emph{blamable}, or the localized connection between the error and what caused the error. For each error, we highlight what (and where) in the \readme{} caused the specific error as a wavy red underline (\figref{fig:linter-overview}C) and \figref{fig:localized-error}. Users can hover over the error to get a description of the identified error. 
A list of all failed lint rules is also presented in the console. 
When available, we use the node position or line number from the markdown AST parse to identify where the lint failed. 
Summative operators (\eg{} \operator{count}, \operator{length}, \operator{threshold}) are attached to specific extracts. For instance, a \operator{count} run on a paragraph extract will be attached to each of the paragraphs in that extract. 
For heuristics like duplicate sentences or sensitive-language flags, when no exact span is given, we map the issue to the first line containing offending text. If no reliable location can be determined, we conservatively default to line 1 so that the console entry and editor remain anchored without implying false precision.

The next rung is \emph{adjustable}. \system{} lint rules can be adjusted by directly making modifications to the rule (such as by increasing the maximum number of allowed emoji), by removing the rule from the set being executed (ignore globally), or by marking a specific line in the \readme{} as being absolved from execution via a comment like 
\verb+<ignore-line-for:RULE-NAME/>+
(ignore locally).

Lastly, is \emph{fixable}, or the ability of the system to automatically fix identified errors. 
For each failing rule, we can generate a potential fix for the identified error via an LLM, which we find to be broadly successful at providing textual fixes. 
Notably, there is an absence of a ``fix all issues'' button. Per \gAgency{}, we want users to be engaged with the process of improving their documentation rather than merely accepting whatever is generated. When fixing individual lint errors, cycles can be introduced wherein fixing one error introduces another. A fix-all (such as in VizLinter~\cite{chen2021vizlinter}) might address this, but preclude users from the opportunity to engage with the design tensions such cycles highlight (reflecting \gAgency{}).  
We note that, just as some CLI-based linters require a \verb+--fix+ flag to engage a fixer, automated fixes are not specifically required unless called for.

\section{Evaluation}

To examine the extent to which \system{} is able to fulfill our design goals, we carried out three studies.
First, we used \system{} as a design probe in a user study to better understand views on documentation linters as mediated by our design goals. 
Second, we compared our approach with naive LLM usage, providing insights into \system{}'s ecological validity.
Finally, we consider expressiveness via application to an alternate domain (recipes), highlighting its applicability to varied communities (\gComm{}).

\subsection{First-use User Study}
\label{sec:user-study}

We conducted a (N=\numParticipants{}) user study in which participants tried out \system{} and used it to reflect on their own perspectives on \readme{} assistance through a semi-structured interview.
This study considers the question: how do users experience documentation linting when it is explicitly designed to support our design goals? (\gAuthorable{}, \gComm{}, \gAgency{})

We recruited \numParticipants{} participants (\pxx{x}) through social media (Bluesky and LinkedIn) and Slack channels, targeting individuals who had at least basic familiarity with \readmes{}, which primarily yielded participants from within our state. 
Each participant was asked to bring their own \readme{} to use during the session. 
Participants had a variety of experiences working with \readmes{} (ranging from somewhat to very familiar), held different roles (5 MS students, 4 PhD students, and 2 industry professionals), and worked in a number of domains (including software engineering, data science,  and machine learning).

The study consisted of three parts. 
First, we asked participants about their experiences with reading and creating \readmes{} to understand their prior background with \readmes{}. 
Then, in the second part, we gave a brief demonstration of \system{} followed by a period of free exploration. 
Then, participants were asked to complete two tasks in the \system{} playground: (1) lint their prepared \readme{} using the rules present in the system, and then (2) create a new rule that they thought would be useful.
In task 1, participants explored \system{} by selecting or customizing linting rules and applying them to their own \readme{}.
Most participants began by browsing the available rules, either individually or by selecting a preset, to understand what each rule checked for. 
After running the linter on the selected rule or rules, they inspected the flagged issues and then decided to accept, reject, or modify the suggested fixes.
During task 2, participants were asked to create a new rule.
Participants relied on the LLM for the initial rule creation, for example creating rules related to contributor lists, image requirements, markdown rendering, or \readme{} scoring. 
In the case where the rules created did not completely fit what the participants expected, they used YAML to edit the syntax, adjust operators, or tailor thresholds to their own document.
Participants did this iteratively in cycles of running the linter, inspecting the output, and then modifying the operators. 
Finally, in the third part, after completion of the tasks, we asked participants to reflect on their experience of using the tool and use it to probe their views about \readmes{} more generally.

To reduce interviewer bias, our semi-structured interviews were conducted using the pair-interview technique \cite{akbaba2023two}.
The interviews were done remotely using Zoom and took around 1 hour.
Participants received a \$40 Amazon gift card as compensation.
Our study was marked exempt by our institution's IRB.  
The interview instrument is in the supplement.

Automatically generated transcriptions were reviewed for correctness and de-identified by the first two authors. 
The first two authors separately open-coded the transcripts, and then compared codes, discussed discrepancies, and iteratively refined a set of shared codes.
After coding the transcripts, the team met several times to group related codes into larger thematic clusters. Following feedback from reviewers, we met again to refine the thematic clusters around the design goals for presentational clarity---echoing the "producing a report" stage of thematic analysis~\cite{braun2006using}

Generally, participant reactions were positive. 
Some participants suggested that it was user-friendly (\pxx{1}, \pxx{5}, \pxx{7}), well laid out (\pxx{4}), and even aesthetically appealing (\pxx{11}).
However, others offered nuance to this position, with several noting a learning curve (\pxx{6}, \pxx{7}, \pxx{8}, \pxx{9}, \pxx{11}).
While these comments are informative, we stress that our probe has not been optimized for usability or learnability, and is instead meant to be a means to explore reactions to this form of tool, which we now describe, organized by design goals.

\subsubsection{Authoring \gAuthorable{}{}}
This goal values users' ability to add or adapt rules to fit their own domain, workflow, or writing style. 
In our study, participants frequently adjusted existing rules, such as by modifying operator configurations.
For example, while testing \lint{detect-hate-speech}, \pxx{11} noted a bug where ``master-slave'' was not flagged, but each term was flagged individually. 
Hypothesizing that the hyphen might be the problem, they attempted to fix the rule by editing the YAML for \operator{count} and \operator{threshold} operators. 
After failing to improve it that way, they pivoted to adjusting the prompt in the \operator{fix-using-llm} operator, which fixed the problem. 
When LLM-generated rules were imperfect, participants also refined or corrected them.
In task 2, \pxx{9} wanted to generate a \readme{} scoring rule to assess their \readme{} quality.
They started by generating a lint rule (via \figref{fig:labeled-rule}B) with the prompt ``\texttt{I would like to get a score for this \readme{} file that how helpful it is for other users}''.
However, this proved to be too vague.
They then edited a \operator{fix-using-llm} operator around which the rule was centered, adding a specific instruction to calculate a score based on section completeness in addition to recommending text changes. Running it again, \pxx{9} was satisfied with its performance. 
The availability of multiple authoring pathways, using the LLM, editing the YAML, or combining both, allowed participants to construct rules in the manner they preferred.

Impeding the authoring process was a learning curve noted by several participants (\pxx{6}, \pxx{7}, \pxx{8}, \pxx{9}, \pxx{11}). Some remarked that understanding \system{} takes time to get used to, particularly editing YAML (\pxx{6}) and understanding the rule generation interface (\pxx{7}).
While such a curve is a barrier to authoring, we suggest that these issues might be addressed through continued design work. For instance, \citet{l2024learnable} found that an unfamiliar DSL could be learned by making use of blended multi-modal interfaces that draw on AI-based tooling. 
This raises additional questions about agency, but it does suggest that this type of obstacle can be addressed.

A facet of authoring is overcoming the blank-page problem. In \system{}, this was achieved through remixing or modifying existing lint rules or by generating new ones from scratch using automated assistance. 
\pxx{4} suggested that \system{}'s value was not merely in the enforcement of standards but rather getting \qt{them into the habit of documenting stuff}, such as by nudging people to write a particular missing section. 
Several appreciated the presets (\pxx{1}, \pxx{2}, \pxx{10})  and the variety of rules (\pxx{1}, \pxx{2}), as it helped them identify usages  they had not been previously aware of.
\pxx{7} explained, \qt{I didn't have a table of contents before, but now I know there is a rule that the \readme{} should have a table of contents; after that, I'll use \system{} to integrate [one]}.
\pxx{6} observed that the number of rules can be overwhelming, which suggests that there is room to improve lint rule presentation such that addressing them is less cognitively burdensome.

Participants envisioned a variety of ways to use \system{}.
Some positioned it as an \qt{end-of-pipeline quality check to identify which part of it is consistent} (\pxx{8}) or as a way to enforce a \qt{common structure of \readme{} files...especially for beginners [that don't] know how to write a \readme{}}. 
Some participants preferred to \qt{change every iteration [for example,] optimizing some parameters} (\pxx{5}, \pxx{8}, \pxx{9}), while others preferred to \qt{[write \readme{} at the end of the project, when everything's done and when it's ready to go public ]}(\pxx{11}).
\pxx{7} and \pxx{9} highlighted its potential for gap-finding, maintaining version accuracy, and quickly validating code-related instructions.
This highlights how the authoring process is varied and personal, suggesting that effective feedback should be available in a variety of different forms and points in time. While \system{}'s playground is primarily focused on a quick iterative feedback loop, the tool can also be used at the end of the writing process to fix issues once all the key pieces are in place.

\subsubsection{Community \gComm{}} A central assumption in our approach, informed by prior work~\cite{crisan2025linting, Cruickshank23SQLFluff}, is that linting is an inherently communal process and that the needs of different communities vary. We find that participants believed this as well and that our design was well matched with that assumption.

In pre-interviews, participants described a good \readme{} as one that includes an introduction of the project (\pxx{3}, \pxx{6}, \pxx{7}, \pxx{8}, \pxx{10}, \pxx{11}) and clear setup steps (\pxx{1}, \pxx{2}, \pxx{3}, \pxx{4}, \pxx{7}, \pxx{10}, \pxx{11}), alongside an explicit structure of the documentation (\pxx{9}, \pxx{11})---such as table of contents and inclusions of specific sections.
We note these observations here as they relate to the structure of our probe.  
They valued documentation that is easy to follow (\pxx{1}), appropriately detailed (\pxx{1}, \pxx{2}), and not overly complicated, avoiding unnecessary information (\pxx{2}).
Additional elements cited were a 
visual demonstration (\pxx{6}, \pxx{11}),
jargon explanation (\pxx{4}), contribution (who developed the project) (\pxx{6}), licensing (\pxx{6}), and so on---echoing \citet{tang2023evaluating} and reaffirming our assumption that \readmes{} interact with multifaceted preferences.

To the end, some expectations of \readme{} properties were domain-specific. 
For example, participants who worked in visualization-oriented projects expect a 
\qt{picture of the visualization design interface overall} 
(\pxx{7}) and sections on \qt{perception, or...results of user study of the visualization you developed} (\pxx{6}).
While machine learning and software engineering participants wanted to know
\qt{what [the] model is doing, what application it is trained for...what data set it is trained [on]...the model architecture...the training steps...[and] the data pre-processing step} (\pxx{9}), and to report performance, with 
\qt{a section that...talks about the accuracy of the model, or the [metrics], the losses and everything about the model} (\pxx{6}).
Evidently, no single quality measure will appropriately capture the needs of every domain, mirroring \gComm{}. By allowing groups to select the lints relevant to them, we  provide targeted and relevant feedback.

Audience needs also shaped the expectation of facilities. 
When sharing \readme{}s outside of their domain, participants valued clear explanations and limited jargon, prioritizing \qt{understanding what explanation would be important in terms of user perspective} (\pxx{9}).
\pxx{5} described cross-domain handoffs in health contexts, noting the value of workflows that a non-specialist can follow: \qt{So let's say I am doing visualizations for someone in [the] health industry...if they provide the workflow [... ensure that] someone who is not from the field [can] understand}. 
By contrast, when \readme{}s target coworkers or future maintainers, participants expected more granular detail, aiming to \qt{fill up the requirements which could be continued...by the future employees} (\pxx{5}) and to be \qt{good enough for the new person to understand the \readme{}...because...it doesn't matter unless the user...knows how to use it} (\pxx{8}).
Even when the audience for a \readme{} is its co-authors, people might forget to act appropriately or courteously. 
\pxx{11} recounted an experience at a hackathon where teammates were
\qt{bombarding the \readme{} file with unnecessary comments and unwanted information....[as well as] incorrect or hateful information and put it in a \readme{} file}, highlighting the value of operationalizing community standards (as linters do) in an easy-to-use way, and underscoring the value of our \gComm{} design goal.  

Finally, participants mentioned collaborative benefits. 
For instance, \pxx{5} recalled that \qt{arguments [about what to put in a \readme{}] are pretty common}, and suggested that \system{} could reduce such debates by externalizing community norms and standards.

\subsubsection{Agency \gAgency{}}
Finally, this goal emphasizes intentional decision-making, where users retain control rather than giving way to automatically provided support. Linters have a well-documented~\cite{chen2021vizlinter, lei2023geolinter} tendency to bias their users into accepting their commentary even if it is inappropriate.
Our participants questioned the advice provided and the structure of the rules, evaluating for themselves if the highlighted changes were appropriate---although, we note, this may change in practice.

A key challenge in this area is interaction with the fixers. 
In both tasks, after running the linter, participants evaluated each flagged issue, choosing which fixes to accept, which to reject, and which to partially apply, often only using the portions they felt were appropriate or aligned with their intentions.
For example, \pxx{5} ran the "Software Libraries" preset on their self-provided \readme{}, which yielded a number of warnings and errors. 
Toggling the fix button for one of the errors, they are able to compare the original text against the proposed fixes. 
Rather than accepting the automated fix, \pxx{5} 
selectively integrated only the desirable edits by copying and pasting the relevant parts. 

In several cases, participants pushed back when fixes were inappropriate or incorrect in their view. 
For example, in task 1, \pxx{9} ran \lint{code-block-consistency} to check for uniform formatting and language usage.
However, the rule flagged errors because the \readme{} contained Python code, while the default configuration only allows Bash.
\pxx{9} resolved this by updating the YAML to include Python as an allowed language.
Running the linter a second time, the rule flagged an incorrect number of indentations within the code blocks.
Believing that the code was visually correct, they modified the pipeline to add an image rendering operator to verify the actual appearance of the output. 
On inspection, they concluded that there was nothing wrong with the indentation and so ignored the error.
Linters inherently risk automation bias, but we found that, at least in some cases, participants did not always treat automated fixes as universally authoritative.

A key way of maintaining control was through the traceability between rules and applied changes. 
Specifically, participants (\pxx{1}, \pxx{3}) valued being able to review and selectively apply edits.
While this functionality is present for some rules, this suggests that a tighter binding of error and source would be useful.
Providing tools that offer feedback that is specific to its cause (rather than other approaches, which might  merely offer a summative quality score) then seems to be a value in retaining user agency. 

While the collection of presets seemed empowering for overcoming the blank page problem, it also carried with it notions of undue authority. 
For instance, \pxx{2} suggested that the presets were useful and chose to rely on them during the study rather than scanning the entire list. This suggests that sets of lints act as mediators of authority.
This echoes a potential line of thinking that sees anything included in the rule set as authoritative and intentional.
While \system{} makes efforts to highlight that rules are malleable, this conceptual bundling may impede that goal, suggesting that additional design work is needed to consistently empower users.

Beyond quality enforcement, participants emphasized time savings, guidance on missing sections, and assistance with maintaining version details (\pxx{3}, \pxx{4}, \pxx{7}, \pxx{9}).
\pxx{3} and \pxx{4} specifically emphasized efficiency, noting that it could guide authors toward meeting documentation standards.
Yet, speeding up some processes or by abnegating agency can mean that key details are overlooked~\cite{mcnutt2025slowness}. 
Providing huge walls of lint failures seems to strongly incentivize use of support tools (\ala{} ESLint's fix flag), suggesting that not providing those may lead to rejection of the tool more generally. 
Balancing these conflicting incentives---quickly addressing annoying feedback, ensuring the document reflects the intended content and form---can be mediated by social structures  (\eg{} by blocking pull requests until they are addressed). 
However, we suggest that designing tools in such a way that this negotiation is simplified is critical.
Our probe leans towards sometimes providing inaccurate fixes to mitigate being annoying (or rather impolite~\cite{whitworth2005polite}), but other approaches may view this tradeoff differently.

\begin{figure*}[t]
    \centering
    \includegraphics[width=\linewidth]{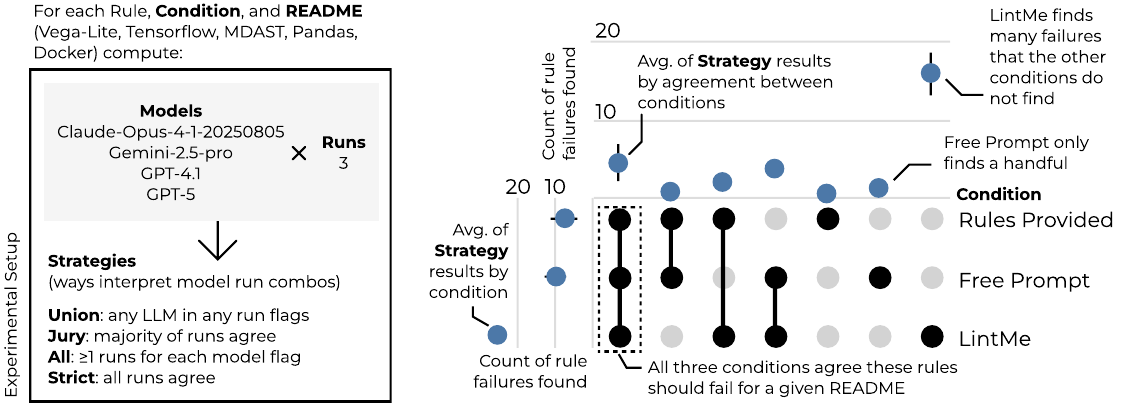}
\caption{\system{} identified more errors across the considered \readmes{} than other conditions, here shown as UpSet plot~\cite{lex2014upset} averaged by strategy. 
Each of identified strategies offers a reasonable approach, and so rather than picking one arbitrarily, we select all of them to highlight the variance in the runs.
    }
    \label{fig:union-upset}
    \Description{A composite figure showing a diagram summarizing the experiment setup and an UpSet plot. The UpSet plot compares the intersections of three experiment conditions, using LLM with rules provided, LLM with free prompt, and LintMe. The middle shows a dot and bar intersection marix overlap for all three experiments, two experiments, and one experiment. Within the different sets, the set with all three experiments is highlighted with description all three agree these rules should fail for a given README.

    The left side of the UpSet plot shows a point chart showing the number of flags given with a confidence interval. LintMe have the most rules flagged, with more than 20. While both LLM flags amount close into 10 with small range of CI.
    
    Above the matrix, a point chart shows the different value for each intersection. The matrix of only LintMe with no intersection, consistenly flags the most rules, followed by the intersection of the three experiments. Across intersections, most sets have smaller counts, with LintMe showing highest values overall.
    }
\end{figure*}

\subsection{LLM Comparison}
\label{sec:llm-study}

\newcommand{\rulesProvided}{\texttt{rules-provided}}
\newcommand{\freePrompt}{\texttt{free-prompt}}

Next, we consider: how does this approach compare with 
using commodity LLMs to improve their \readmes{}? 
To answer this question, we conducted a small experiment comparing lint rules executed using LLMs and \system{}.
We selected a set of varied \readmes{}: Vega-Lite (a visualization library), Tensorflow (a machine learning library), MDAST (a markdown AST specification), Pandas (a data analysis library), and Docker (a containerization tool). 
Vega-Lite's \readme{} features images, nested sections, and examples, while TensorFlow's is a simpler, but text-dense, structure that includes dates and version badges.
MDAST's is a lengthy, domain-specific specification with formal definitions and extensive code blocks. 
Pandas's is concise and example-oriented with code snippets and a clear structure.
Docker's reflects a command-heavy style with numerous terminal code blocks but minimal narrative explanation.

We evaluated each \readme{} using four LLM models (\figref{fig:union-upset}) under two prompting conditions, \rulesProvided{} (which included the same list of rules as was available in \system{} at the time of evaluation) and \freePrompt{} (which included no specific suggestions on what might be wrong).
This approach was designed to replicate how a contemporary developer might give a \readme{} to an LLM and ask it to improve it---which participants in our user study noted as being a common usage.
Prompts (available in the appendix) were relatively short (just 7-8 lines excluding inputs) and did not make use of our documentation beyond the natural language list of rules. 
Model temperature was set to 0 where supported (GPT-5 does not support temperature).

\begin{figure*}[t]
    \centering
    \includegraphics[width=\linewidth]{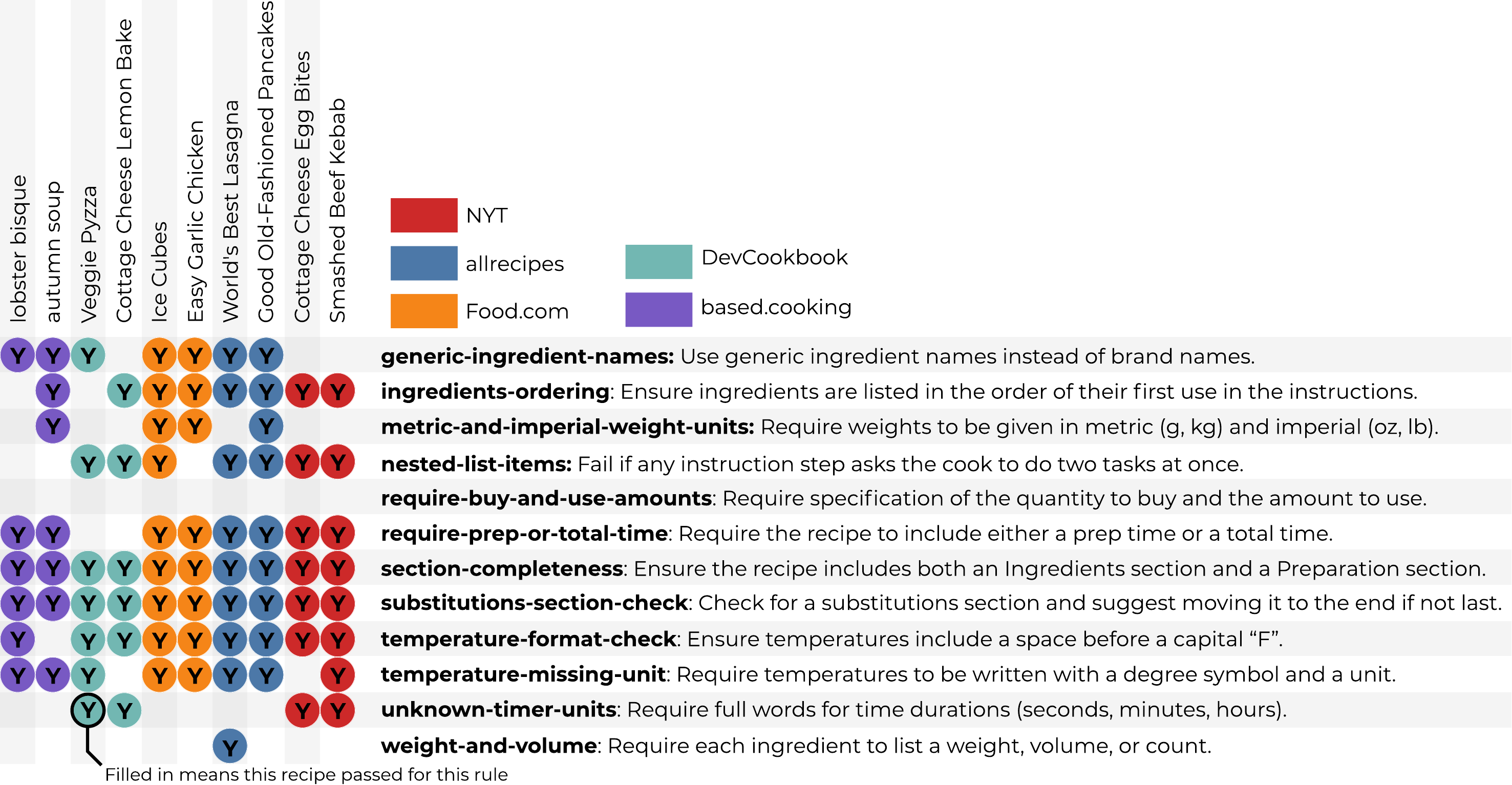}
    \caption{To explore the expressivity of \system{}, we generated a collection of rules that evaluate  desirable style or substance properties, such as not using multi-part instructions within a single step. With a small amount of format coercion, \system{} can evaluate real-world recipes. This heatmap-style table shows whether or not a given recipe triggers a lint rule, for instance, the highlighted cell indicates that \lint{weight-and-volume} passed for the ``Veggie Pyzza'' recipe. }
    \label{fig:recipe-line}
    \Description{
A binary heatmap show whether or not a set recipes trigger a lint rule. The columns are recipes, such as lobster bisque. The rows are lint rules, such as "nested-list-items: Fail if any instruction step asks the cook to do two tasks at once". A cell is present if the rule marks the recipe as passing. About a quarter of the cells seem to be failing.
    }
\end{figure*}

Each model-prompt-dataset combination was run three times to explore  model variability, generating a set of issues for a given \readme{}, which we viewed as failing lint rules.
To evaluate these runs, we mapped each of the identified issues to a predefined set of \system{} rules.
One author conducted an initial mapping and then met with another to discuss and check it. 
This yielded three sets per dataset: Free Prompt (LLM), Rules Provided (LLM), and \system{}.

Across all five datasets, we found that \system{} consistently flagged a greater number of problems than the LLMs---a finding we summarize in \figref{fig:union-upset}. 
Most rules flagged by the LLMs were also flagged by \system{}, with \system{} typically finding several additional issues.
Some rules were reliably detected by all systems across datasets, such as \lint{table-of-contents} and \lint{section-completeness}.
\system{} identified several issues that none of the LLMs noted, including \lint{no-hate-words} (\eg{} avoiding terms like cripple), \lint{use-objective-language} (information is factual, unbiased, and clearly presented, avoiding promotional or vague language), and \lint{availability} (is the \readme{} in a standard location?), highlighting the value of \system{}'s enumeration of specific standards.

To some degree, this finding is unsurprising. Many of \system{'s} rules use LLMs, which have been better tuned than the naive usage employed in this experiment---although many rules do more than merely call LLMs.
Aligned with this observation, we highlight that \system{}'s granular itemized approach flags ($\mu=25.4$ issues on average across datasets and runs) more issues than \freePrompt{} ($\mu=9.6$) or \rulesProvided{} ($\mu=7.25$).
We found the GPT-5 had the most commentary that was most consistently aligned with that of \system{}---although \system{} consistently identified more violations. 
This performance may be attributable to its different model architecture, marketed as a reasoning model.

While \system{}  identified more errors than the compared models, \rulesProvided{} sometimes approached a comparable level of granularity.  
However, this is not to say that a linter can be simply replaced by an appropriate LLM call. 
Linters are more than merely noting that there is an error---there are several additional rungs in the linting ladder. 
\freePrompt{} is not \emph{adjustable} or \emph{blamable}. 
\rulesProvided{} is modestly \emph{adjustable} (specific rules can be removed, akin to our global ignore), but specific lines will be difficult to adjust.
It is possible that the violations identified by these LLMs could be used to instrument fixes, however, this rung was not explored in this evaluation.

Moreover, part of the value of a linter is the curation of a list of rules, which form community standards (\gComm{}). While these can always be materialized, the labor of building those collections is a reflection of their value.

\subsection{Expressivity Case Study}
\label{label:recipe-study}

Finally, we evaluate the expressiveness of our approach. 
Beyond \readmes{} there are numerous forms of structured or semi-structured textual communication with associated cultural standards or mores  that it may be beneficial to lint. 
Sometimes these rules are implicit. For instance, instruction manuals for board games often follow an order of discussing the pieces, the setup, the rules, and sometimes include components such as turn summaries or examples of play---the inclusion or exclusion may lead those rules to be more or less effective~\cite{riggleman2024technical}.
Sometimes guidelines are explicit,  as in style guides or lists of best practices---such as the ``Ten simple rules for'' series in PLoS journals ~\cite{plos10SimpleRules}.
Just as with documentation, other domains often have collections of bad practices which are easy to accidentally do---such as forgetting to include alternative text.

Here, we focus on the prominent case of culinary recipes. 
Recipes come in a great many varieties and levels of detail.
In some venues, it may be acceptable to use heuristics like ``some'' of an ingredient, while in others specific measurements might be required.
Recipes make an ideal candidate for this form of evaluation, because they can typically be coerced into a markdown file without loss of functionality, and there are explicit style guides as well as implicit practices.
Among the style guides we identified online~\cite{missouri-recipe-style-guide, vt-fst384-how-to-write-recipe, gillingham-2023-how-to-write-recipe}, there were a variety of rule types and forms. 
For instance, ``For temperatures, use a capital ``F'' with one space after the numerals'',
``Use common terms for cookware and cooking utensils'' (\eg{} sheet pan (not jelly roll pan)),
``Substitutions for ingredients can be included at
the end of the recipe'',
and ``Use generic names of ingredients (semi-sweet chocolate chips, not ``Tollhouse chips'')''.
These requirements vary widely in degree of specificity, highlighting an opportunity to explore \system{}'s expressivity. 

Based on both published style guides and informal community opinions, we constructed a set of twelve recipe-specific lint rules and bundled them into a preset called ``Recipe Rules''. 
To evaluate this set of rules, we sampled recipes from several high-visibility cookery sources (AllRecipes~\cite{nyt-smashed-beef-kebab, nyt-cottage-cheese-egg-bites}, The New York Times Cooking~\cite{allrecipes-good-old-fashioned-pancakes, allrecipes-worlds-best-lasagna}, and Food.com~\cite{foodcom-easy-garlic-chicken, foodcom-ice-cubes}) and two open-source repositories of markdown-formatted recipes on GitHub (DevCookBook~\cite{devcookbook-cottage-cheese-lemon-bake, devcookbook-veggie-pyzza} and based cooking~\cite{basedcooking-autumn-soup, basedcooking-lobster-bisque}). 
For sites where recipes were provided as plain text or HTML, we used an LLM (GPT-5) with a conservative prompt (``convert this to markdown without changing content'') to obtain markdown text suitable for linting. 
This conversion step highlights a potential direction for expanding \system{} beyond markdown-first domains. 

Each recipe was evaluated against the 12 rules, which were expressed using existing operators originally designed for \readme{} linting, namely \operator{customCode}, \operator{evaluateUsingLLM}, \operator{extract}, \operator{regexMatch}, \operator{count}, and \operator{threshold}. 
The details of the rule implementation can be found in the playground. 
The composability of these operators demonstrates that the same primitives can be appropriately used across domains with very different conventions. 
Each rule was manually spot checked for accuracy. 
Results (summarized in \autoref{fig:recipe-line}) show that even widely used recipe sites routinely violate basic style guidance. 
Common issues included omission of degree units in temperatures and deeply nested procedural steps. 
While temperature format might be evaluated using prior markdown linting tools, \lint{generic-ingredient-names} would be challenging to implement without a comprehensive list of every ingredient and brand---highlighting the value of our mixed programmatic and LLM evaluation strategy.  
Interestingly, within a given website, violations tended to recur systematically, suggesting the influence of editorial standards or templates.

From these results, we are optimistic that this strategy can usefully automate the style guide in a variety of additional domains. 
This brief sketch misses some of the higher-level semantics that are embedded in recipes. 
For instance, it would not catch errors related to execution order (\eg{} putting a breakfast casserole in the oven before the eggs have been added). 
Future work on evaluating recipes might draw on \citeetals{fan2023understanding} system for extracting flow graphs from recipes for analysis. 
More broadly, however, we emphasize that this suggests there is a wide range of additional domains to which \system{}-style evaluation might be applied (\gAgency{}). 

\section{Limitations}
\label{sec:limitations}

As with any work, our system and studies had a variety of limitations. 
\system{} is a design probe and so some engineering decisions made in the interest of exploring what is possible---limiting its applicability in its current form. 
For instance, the \operator{customCode} and \operator{execute} operators run arbitrary JavaScript and shell commands, meaning they may be vulnerable to malicious use. 
Our demo sandbox is run within a single isolated virtual machine, so the range of effect of malicious code is limited to the shared database of rules.
However, \system{} can be run as a CLI tool, where the risks are similar to running any arbitrary JavaScript package and necessitate similaring vetting of rules. 
Similarly, our \operator{EvaluateUsingLLM} raises potential privacy issues for \readmes{} not meant to be shared with the public (such as in documentation internal to companies), which was noted by participants in our user study.

To that end, many of our more complex lint rules make use of LLMs as a backing implementation. 
For instance, \lint{neutral-tone} asks an LLM to evaluate whether or not the sought property is met. 
While in initial testing, these checks worked appropriately, LLMs' non-determinism and propensity to hallucinate means that they will sometimes be wrong (\ie{} give false positives or negatives) or may not appreciate the full scope or complexity of a particular task.
While these are important concerns, we forwent addressing them in favor of using them as scaffolding for exploration, which is itself a limitation of our design probe approach. 
For instance, in the domain of code, we can employ complex analyses to ascertain if particular properties hold, but this is not necessarily the case for every possible lint of interest. 
As we noted in our expressiveness case study, our ability to analyze the semantics of a recipe is limited, but can be improved through developing additional analysis lenses. 
While our approach can approximate or attempt any lint that might be desired, the resulting lint may not accurately capture the intended behavior. For instance, in cases with substantial subtlety (such as legal documents~\cite{han2020textlets}) or logical semantics that cannot be captured through natural language (as in our recipe example). 
Future work might explore linter judgments other than right or wrong, such as unsure or insufficient information.

Next, we note that our studies have several limitations.
Our user study included participants with a variety of backgrounds; however other communities and \readme{} users may view our tool differently. Future work might explore how specific open-source communities approach \readmes{} with multiple stakeholders and the role that automated assistance might be provide therein.
Due to an administrative error, one participant who is employed by one of the team members was interviewed in the user study. This team member joined after the employee was interviewed and so we believe the bias towards this system is minimal---although we do not take the comments of any of the participants as objective usability evaluations. 
Similarly, our LLM comparison study had several limitations. 
One complicating factor with analyzing these results is that there is not a single ground truth for many of these rules. 
For instance, who decides if a tone of writing is sufficiently neutral? 
Instead, we stress that \system{} simply surfaces \emph{more} information. 
While we believe from our own experience that the answers provided are right, such an evaluation is biased by our own perspectives.  
Finally, the number of \readmes{} evaluated was small, and the results may shift with additional prompt engineering.
This evaluation was a short check to evaluate whether naive usage could easily replace \system{}, rather than a systematic evaluation of the comparisons between these systems.

\section{Discussion}
\label{sec:disco}

There is more to \readmes{} than mere stylistics. They are multifaceted documents~\cite{tang2023evaluating}, containing a range of different forms of content that are often specific to expectations of particular domains or groups. 
Through this work, we demonstrate that a linter can effectively identify deviations from non-trivial communal standards (\gComm{}). 
Our design probe, \system{}, demonstrates this by interweaving programmatic execution with LLM support, enabling end-user authorable rules (\gAuthorable{}).
Using the probe, we explored the space of possible documentation properties that can be evaluated with modern methods and interface designs.
We find that a linter can offer a powerful and flexible foundation for documentation analysis, that, unlike previous linters~\cite{proselint, alex, writegood, markdownlint_node, markdownlint_ruby, remarklint}, is able to evaluate both style and substance, including evaluation of more amorphous properties such as tone, structure, and comprehensibility---affording users a substantially greater ability to evaluate their properties of interest (\gAgency{}).

Moreover, this work looks to the potential of having an automated assistant that can offer structured, granular feedback for \readmes{}. 
We explored how this work might be applied to complex tasks to evaluate rules (such as whether or not code will run) and additional domains (recipes). 
While this approach inherits the limitations of LLMs that drive some of its analysis, hence precluding us from analyzing some complex or nuanced properties, we found that we were able to reasonably implement rules for a wide range of common conventions or standards.

In doing so, we noted a variety of tensions and trades-offs inherent to this approach---for instance, the role of automated fixers as both domineering entities and affordances for dealing with large collections of errors. 
As with any tool that mediates sociotechnical standards, there is no single solution that will deal with all requirements and edge cases. We suggest that the linter design offers an enticing structure for encoding collective judgment in documentation, sensitive to both human taste and local standards, although there are tradeoffs.
We conclude by continuing to reflect on open questions prompted by our probe.

\subsubsection{Linter System Design}
A crucial consideration in this design is the decision to author lint rules through a DSL. 
This decision was made following prior work~\cite{vale, mcnutt2024mixing} and was done to center \gComm{} and \gAuthorable{}. 
While participants in our study were able to use our DSL, they observed that the YAML representation carried a non-trivial learning curve---echoing prior works~\cite{mcnutt2021integrated, l2024learnable}. 
However, we found experientially and in our user study that the Generate Rule feature (\figref{fig:labeled-rule}B) frequently produced rules that were ``good enough'' to serve as starting points.

This suggests that our chosen operator granularity appropriately balances usability and customizability, however, other sizes were possible. For instance, we might have followed vale~\cite{vale} and used larger, pre-structured rules that were simpler to tune but harder to compose. At the other end of the spectrum smaller operators might have made rule creation richer and more composable~\cite{horowitz2023engraft}, but added significant complexity and learning barriers. 
While there is single best approach across this granularity spectrum, based on our user study where participants were able to successfully tweak rules to taste, the one used here seemed to be appropriate for our use case. Use cases in other domains or with other users might prefer a different size. No one grammar design will rule them all~\cite{mcnutt2022no}, as there is no single universal community or context.

Next, another possible approach might have involved using LLMs instead of using a DSL, echoing what might be called \emph{vibe linters}, such as GPTLint~\cite{fischer2024gptlint}.
Our LLM Comparison study suggests that our approach was more precise compared to commodity LLMs, it is not clear that this will always be the case; it is possible that LLMs could become more expressive and more accurate. 
Critically, we suggest that DSLs offer a segmentation of intent that is difficult to capture purely through natural language. In addition to providing conceptual building blocks for understanding what a rule should do, such segmentation offers \emph{common objects}~\cite{beaudouin2021generative} through which collaborators can reason about rule provenance, provide accountability for rules made and ignored, and concretely negotiate the terms of what is expected. While LLM-mediated approaches like AI-chains~\cite{wu2022ai} offer aspects of this, the uncertainty latent to such specification impedes some of these benefits.

\subsubsection{Identified and unresolved issues} 
Through this probe, we identified some  usability questions about linters for which we do not have ready answers. 
For instance, providing the user a large list of errors to deal with can be overwhelming, potentially leading to them being ignored---akin to the experience of having too many \LaTeX{} errors.
\system{} attempts to address this by attaching the errors to specific portions of the markdown file that caused those errors; however if there are too many rules, this may still be overwhelming---which was noted by our study participants. 
Future work could explore automatic hierarchical bundling that could guide the user through fixes or automatic tagging of rules so that users can focus on individual dimensions of authoring one at a time, depending on their pre-rated severity (\eg{} gradations of warnings versus errors).

Another key issue in linters is automation bias, or the human tendency to give undue trust or value to automatically provided suggestions. 
As with any system that automatically surfaces knowledge, linters are particularly susceptible to automation bias~\cite{lei2023geolinter, chen2021vizlinter, mcnutt2023study}.
However, this bias can be magnified or reduced by design decisions.
For instance, \citet{chen2021vizlinter} observe that fixing one rule failure (in a visualization) may uncover another, which may in turn uncover yet another, and so on. 
They develop a constraint solver-based fixer that resolves all issues.
While this is effective, this strategy is strongly prone to automation bias.
In analogy, code formatters like prettier~\cite{prettier} automatically fix issues like white-spacing, line breaks, and so on, which are of limited importance to the task of writing code.
For domains where human intervention is not meaningful, highly automated fixers seem useful.
However, for domains where human interaction is important, having the fixing actions be somewhat  frictionful~\cite{mcnutt2025slowness} may usefully provide opportunities for creativity or reflection.
While it likely varies by domain, future work might investigate which parts are worth human attention and which should be automated.

Lastly, the effect, in practice, of documentation linters or our customizable approach to linters is not yet known. 
Based on the wealth of different linters available in different ecosystems~\cite{awesome_lint}, this form of customization and rule building seems appropriate, particularly in light of the explicitly social nature of linters (as highlighted in the SQLFluff~\cite{Cruickshank23SQLFluff} docs).
However, future work might usefully explore this in practice. For instance, an empirical review of documentation conflicts in open source projects or an interview study with open-source maintainers about the role of linters and documentation might reveal additional nuances, particularly in areas with multiple stakeholders or use cases. For instance, a \readme{} might be used by developers to keep deployment processes consistent, by new users to identify if a library is useful, and as a shared community artifact to crystallize values. As \citet{yang2024considering} point out, documentation often must be all of these things and more. Navigating these contrasting views is not straightforward and requires careful alignment between intended community and artifact.

\subsubsection{Towards broader documentation linting}
While documentation takes a broad range of forms, this work only focuses on markdown files containing \readmes{}.
Our expressiveness evaluation demonstrated that our approach can be directly applied to  another domain manifestable as a single markdown file, \ie{} culinary recipes. 
Documentation often comes in a much wider array of forms, often involving websites with many pages.
In future work, we intend to expand the scope of our linter to capture other documentation formats (such as Markdown, AsciiDoc, reStructuredText, as vale~\cite{vale} already does), as well as full documentation websites, such as those generated by documentation systems such as Sphinx---either on a code level, the generated documentation, or a mixture thereof.

Beyond analysis of the content, there are also challenging factors to evaluate, such as the temporal or conceptual distance between two concepts---for instance, traversing between parts of the documentation that are closely related but require substantial clicking and scrolling to get between, adds friction to the process of using that documentation.
Another challenge is that developers tend not to trust automatic documentation generators, as observed by \citet{wu2021exploring}. 
Integrating a documentation linter into such a process might give sufficient opportunities for human-in-the-loop clarification of issues to restore trust.
Similarly, future work on our linter could add arbitrary additional sources of data, allowing lints to evaluate a variety of different sources of truth in concert. For instance, a lint validating install instructions might consult a CI pipeline, test suite, or program analysis system (echoing the technique employed by \citet{zhou2018automatic}) as a means to increase trust in both documentation and evaluations of it.

Complicating straightforward evaluation of markdown systems like Sphinx, is that some documentation systems are automatically generated from the code they document. 
This is beneficial in that it keeps the documentation up to date, but it may make analysis of that documentation more challenging. 
More complicated still are the changes to the code, which may necessitate significantly altered descriptions of the code (for instance, if an essential abstraction has been radically altered or added).
Ideally, a holistic documentation evaluation system should be able to identify and surface such situations for human attention.

\subsubsection{Curating and applying knowledge bases}
Developing and extracting domain knowledge to supply to a linter is challenging. 
The collection of lint rules that make up \system{} is based on the evaluation and tabulation of a large number of prior works.
Similarly, \citeetals{mcnutt2024mixing} color palette linter required a large survey of extant guidelines and domain knowledge. 
The core knowledge base in Draco~\cite{yang2023draco} draws on a close reading of prior visualization theory on expressiveness and effectiveness.
If linters are to be for communities, then a key challenge is making it straightforward to develop knowledge bases.
Knowledge bases are valuable, for linters and for documentation, among many other tasks,  and so developing tools to help users build and verify those knowledge bases is useful future work. 
This might take the form of authoring developer tools, such as a block-based editor, as in varv~\cite{borowski2022varv}. 
Data Wave's~\cite{elvira2024datawave} use of a visual analytics-style approach is also compelling. 
Another strategy might involve a programming-by-demonstration style interface wherein the user provides examples and the system generates a rule checking for those examples.  
In any case, future work should prioritize lowering the technical barriers to knowledge base construction, enabling domain experts to directly contribute their expertise without requiring deep technical implementation skills.

\begin{acks}
We thank the National Science Foundation for supporting this work (III-2402719). 
We appreciate our reviewers feedback and careful commentary.
We are grateful to Katy Koenig, Shiyi He, and the Isaacs lab for their support in various ways, including providing commentary, complaints, feedback, and piloting. 
We thank Joseph Kato for his valuable post-publication corrections.
Finally, we are deeply appreciative of our study participants.
\end{acks}

\bibliographystyle{ACM-Reference-Format}
\bibliography{bib}

\appendix

\input{appendix}

\end{document}

%% file: preamble.tex
\newcommand{\ie}{{i.e.,}}
\newcommand{\eg}{{e.g.,}}

\newcommand{\ala}{{\`a la}}

\newcommand{\readme}{README}
\newcommand{\readmes}{READMEs}

\newcommand{\system}{LintMe}

\newcommand{\secref}[1]{\hyperref[#1]{Sec.~\ref*{#1}}}
\newcommand{\appendixref}[1]{\hyperref[#1]{Appendix~\ref*{#1}}}
\newcommand{\figref}[1]{\hyperref[#1]{Fig.~\ref*{#1}}}
\newcommand{\eqnref}[1]{\hyperref[#1]{Eqn.~\ref*{#1}}}
\newcommand{\tabref}[1]{\hyperref[#1]{Table ~\ref*{#1}}}

\newcommand{\citeetals}[1]{\citeauthor{#1}'s~\cite{#1}}

\newcommand{\goal}[1]{DG$_{\text{\emph{#1}}}$}
\newcommand{\gComm}[1]{\goal{comm}}
\newcommand{\gAuthorable}[1]{\goal{author}}
\newcommand{\gAgency}[1]{\goal{agency}}
\newcommand{\gLadder}[1]{\goal{ladder}}

\newcommand{\operator}[1]{\textsc{#1}}
\newcommand{\lint}[1]{\ul{\texttt{#1}}}
\newcommand{\lintwithoutul}[1]{\texttt{#1}}  
\newcommand{\numParticipants}{11}

\def\subsubsec#1
{\subsubsection{#1}}

\newlength{\roundtagcontent}
\newlength{\roundtagwidth}
\newcommand{\roundTag}[3]{%
  \begingroup
  \setlength{\fboxsep}{1.2pt}%
  \settowidth{\roundtagcontent}{\strut\hspace{2pt}#3\hspace{2pt}}%
  \setlength{\roundtagwidth}{#1}%
  \ifdim\roundtagcontent>\roundtagwidth\relax
    \setlength{\roundtagwidth}{\roundtagcontent}%
  \fi
  \colorbox{#2!35}{\makebox[\roundtagwidth][c]{\strut\hspace{2pt}#3\hspace{2pt}}}%
  \endgroup
}

\definecolor{lred}{HTML}{E57373}
\definecolor{lgreen}{HTML}{81C784}
\definecolor{lblue}{HTML}{64B5F6}
\definecolor{lorange}{HTML}{FFB74D}
\definecolor{lpurple}{HTML}{BA68C8}
\definecolor{lgray}{HTML}{B0BEC5}

\newcommand*\rot{\rotatebox{90}}

\usepackage{soul}
\usepackage{url}

\definecolor{linkColor}{HTML}{257E98}
\newcommand\asLink[2]{\textcolor{linkColor}{\href{#1}{\ul{#2}}}}

\newcommand{\systemURL}{\asLink{https://lintme.netlify.app/}{lintme.netlify.app}}

\newcommand{\osf}{
\asLink{https://osf.io/sbfve/}{osf.io/sbfve/}
}

\newcommand{\github}{\asLink{https://github.com/HimaMynampaty/lintme}{github.com/HimaMynampaty/lintme}}

\definecolor{quoteColor}{HTML}{DAB1DA}
\newcommand\qt[1]{\emph{``#1''}}

\newcommand\pxx[1]{\textbf{P}$_{\text{#1}}$}

%% file: abstract.tex
\readmes{} shape first impressions of software projects, yet what constitutes a good \readme{} varies across audiences and contexts.
Research software needs reproducibility details, while open-source libraries might prioritize quick-start guides. 
Through a design probe, \system{}, we explore how \emph{linting} can be used to improve \readmes{} given these diverse contexts, aiding style and content issues while preserving authorial agency. 
Users create context-specific checks using a lightweight DSL that uses a novel combination of programmatic operations (\eg{} for broken links) with LLM-based content evaluation (\eg{} for detecting jargon), yielding checks that would be challenging for prior linters.
Through a user study (N=11), comparison with naive LLM usage, and an extensibility case study, we find that our design is approachable, flexible, and well matched with the needs of this domain. 
This work opens the door for linting more complex documentation and other culturally mediated text-based documents.

%% file: figures/comparision-table-revisited.tex
\definecolor{lred}{HTML}{E57373}
\definecolor{lgreen}{HTML}{81C784}
\definecolor{lorange}{HTML}{FFB74D}

\newcommand{\lBox}[2]{\colorbox{#1!35}{\strut\textsf{#2}}}
\newcommand{\lBoxSmall}[2]{\colorbox{#1!35}{\strut\scriptsize\textsf{#2}}}

\newcommand{\no}{\lBoxSmall{lred}{N}}
\newcommand{\maybe}{\lBoxSmall{lorange}{P}}
\newcommand{\yes}{\lBoxSmall{lgreen}{Y}}

\begin{figure*}[t]

    \small
    \centering
    \footnotesize
    \setlength{\tabcolsep}{0.1pt}
    \setlength{\fboxsep}{4pt} 
    {
        \begin{tabular}{rcccccccccccccccccccccccccccc}
Name\hspace{0.075in} & 
\rot{\lint{execution}} & 
\rot{\lint{code-formatting}} & 
\rot{\lint{link-formatting}} & 
\rot{\lint{link-validation}} & 
\rot{\lint{formatting-and-styling}} & 
\rot{\lint{style-consistency}} & 
\rot{\lint{structure-style}} & 
\rot{\lint{structure-consistency}} & 
\rot{\lint{terminology}} & 
\rot{\lint{sensitive-content}} & 
\rot{\lint{sections completeness}} & 
\rot{\lint{repetition}} & 
\rot{\lint{sentence-length}} &  
\rot{\lint{TOC}} & 
\rot{\lint{no-jargon}} & 
\rot{\lint{visual-clarity}} & 
\rot{\lint{alt-text}} & 
\rot{\lint{cross-platform}} & 
\rot{\lint{availability}} & 
\rot{\lint{objectivity}} & 
\rot{\lint{hate-words}} & 
\rot{\lint{spelling-and-grammar}} & 
\rot{\lint{content-variety}} & 
\rot{\lint{date-related}} &  
\rot{\lint{end-with-new-line}} & 
\rot{\lint{emoji-Lint}}\\
\midrule  

& \multicolumn{4}{c}{\roundTag{1.4cm}{lpurple}{Accuracy}}
& \multicolumn{5}{c}{\roundTag{2.0cm}{lpurple}{Consistency}}
& \multicolumn{1}{c}{\roundTag{0.35cm}{lpurple}{S}}
& \multicolumn{3}{c}{\roundTag{1.05cm}{lpurple}{Concise}}
& \multicolumn{5}{c}{\roundTag{2.0cm}{lpurple}{Accessibility}}
& \multicolumn{1}{c}{\roundTag{0.35cm}{lpurple}{V}}
& \multicolumn{1}{c}{\roundTag{0.35cm}{lpurple}{O}}
& \multicolumn{1}{c}{\roundTag{0.35cm}{lpurple}{R}}
& \multicolumn{1}{c}{\roundTag{0.35cm}{lpurple}{Q}}
& \multicolumn{1}{c}{\roundTag{0.35cm}{lpurple}{A}}
& \multicolumn{3}{c}{\roundTag{1.05cm}{lpurple}{Others}} \\


\midrule  

markdownlint~\cite{markdownlint_ruby} \hspace{1mm} & \no{} & \yes{} & \yes{} & \no{} & \yes{} & \yes{} & \yes{} & \maybe{} & \maybe{} & \maybe{} & \no{} & \maybe{} & \yes{} & \yes{} & \maybe{} & \no{} & \yes{} & \no{} & \no{} & \maybe{} & \maybe{} & \no{} & \maybe{} & \yes{} & \yes{} & \no{} \\
DavidAnson markdownlint~\cite{markdownlint_node} \hspace{1mm} & \no{} & \yes{} & \yes{} & \no{} & \yes{} & \yes{} & \yes{} & \maybe{} & \maybe{} & \maybe{} & \yes{} & \maybe{} & \yes{} & \yes{} & \maybe{} & \no{} & \yes{} & \no{} & \no{} & \maybe{} & \maybe{} & \maybe{} & \yes{} & \yes{} & \yes{} & \maybe{} \\
mdformat~\cite{mdformat} \hspace{1mm} & \no{} & \yes{} & \yes{} & \no{} & \yes{} & \yes{} & \maybe{} & \no{} & \no{} & \no{} & \no{} & \maybe{} & \no{} & \maybe{} & \no{} & \no{} & \no{} & \maybe{} & \no{} & \maybe{} & \no{} & \no{} & \no{} & \no{} & \yes{} & \no{} \\
remark-lint~\cite{remarklint} \hspace{1mm} & \no{} & \yes{} & \yes{} & \yes{} & \yes{} & \yes{} & \yes{} & \yes{} & \maybe{} & \maybe{} & \yes{} & \maybe{} & \maybe{} & \yes{} & \maybe{} & \no{} & \maybe{} & \maybe{} & \yes{} & \maybe{} & \maybe{} & \yes{} & \maybe{} & \maybe{} & \yes{} & \yes{} \\
textlint~\cite{textlint} \hspace{1mm} & \no{} & \no{} & \yes{} & \yes{} & \yes{} & \no{} & \no{} & \yes{} & \yes{} & \maybe{} & \yes{} & \maybe{} & \yes{} & \no{} & \maybe{} & \no{} & \maybe{} & \maybe{} & \no{} & \maybe{} & \yes{} & \yes{} & \no{} & \maybe{} & \yes{} & \maybe{} \\
proselint~\cite{proselint} \hspace{1mm} & \no{} & \no{} & \no{} & \maybe{} & \no{} & \maybe{} & \no{} & \no{} & \yes{} & \yes{} & \no{} & \maybe{} & \no{} & \no{} & \yes{} & \no{} & \no{} & \no{} & \no{} & \maybe{} & \yes{} & \maybe{} & \no{} & \maybe{} & \no{} & \no{} \\
vale~\cite{vale} \hspace{1mm} & \no{} & \yes{} & \yes{} & \maybe{} & \yes{} & \maybe{} & \yes{} & \yes{} & \yes{} & \yes{} & \yes{} & \yes{} & \yes{} & \yes{} & \yes{} & \no{} & \yes{} & \no{} & \no{} & \maybe{} & \yes{} & \yes{} & \yes{} & \yes{} & \yes{} & \yes{} \\
content-linter~\cite{github_content_linter} \hspace{1mm} & \no{} & \yes{} & \yes{} & \maybe{} & \yes{} & \yes{} & \maybe{} & \maybe{} & \maybe{} & \no{} & \maybe{} & \no{} & \no{} & \no{} & \no{} & \no{} & \yes{} & \no{} & \no{} & \no{} & \no{} & \no{} & \no{} & \no{} & \no{} & \no{} \\
alex~\cite{alex} \hspace{1mm} & \no{} & \no{} & \no{} & \no{} & \no{} & \no{} & \no{} & \no{} & \no{} & \no{} & \no{} & \no{} & \no{} & \no{} & \maybe{} & \no{} & \no{} & \no{} & \no{} & \no{} & \yes{} & \no{} & \no{} & \no{} & \no{} & \maybe{} \\
write-good~\cite{writegood} \hspace{1mm} & \no{} & \no{} & \no{} & \no{} & \no{} & \no{} & \no{} & \no{} & \no{} & \no{} & \no{} & \maybe{} & \no{} & \no{} & \no{} & \no{} & \no{} & \no{} & \no{} & \maybe{} & \maybe{} & \no{} & \no{} & \no{} & \no{} & \no{} \\
standard-readme~\cite{standardreadme} \hspace{1mm} & \no{} & \no{} & \no{} & \no{} & \maybe{} & \no{} & \yes{} & \yes{} & \no{} & \no{} & \yes{} & \no{} & \no{} & \yes{} & \no{} & \no{} & \no{} & \no{} & \no{} & \no{} & \no{} & \no{} & \maybe{} & \no{} & \no{} & \no{} \\
awesome-lint~\cite{awesome_lint} \hspace{1mm} & \no{} & \yes{} & \yes{} & \yes{} & \yes{} & \yes{} & \yes{} & \yes{} & \maybe{} & \no{} & \yes{} & \no{} & \no{} & \yes{} & \no{} & \no{} & \no{} & \no{} & \yes{} & \no{} & \no{} & \yes{} & \no{} & \no{} & \yes{} & \no{} \\
grammarly~\cite{grammarly_website} \hspace{1mm} & \no{} & \no{} & \no{} & \no{} & \no{} & \no{} & \no{} & \no{} & \maybe{} & \no{} & \no{} & \maybe{} & \yes{} & \no{} & \yes{} & \no{} & \no{} & \no{} & \no{} & \no{} & \yes{} & \yes{} & \no{} & \yes{} & \no{} & \no{}\\
 \bottomrule
        \end{tabular}}\\[1em]
        
            \textbf{Y} = Supported, \textbf{N} = Not Supported, \textbf{P} = Partially Supported. If a tool supports custom rules (such as via plugins), if implementation is straightforward it is marked \textbf{Y} and \textbf{P} otherwise. \\
            \roundTag{0.35cm}{lpurple}{S} = Security,
            \roundTag{0.35cm}{lpurple}{V} = Availability,
            \roundTag{0.35cm}{lpurple}{O} = Objectivity,
            \roundTag{0.35cm}{lpurple}{R} = Reputation,
            \roundTag{0.35cm}{lpurple}{Q} = Quality,
            \roundTag{0.35cm}{lpurple}{A} = Appearance
    \caption{
    Prior tools have explored different approaches to linting markdown files in general and in \readmes{} specifically. 
    However, most tools have focused on either markdown style (\eg{} correct markdown usage) or on natural language style (\eg{} avoiding hateful terms). 
    Here we draw a sample of representative lints from across \citeetals{tang2023evaluating} documentation typology (denoted in  \roundTag{0.35cm}{lpurple}{purple}) for a variety of linters. While there are other linters and other lints, this collection offers a representative sample of the type of functionality available. 
    Lastly, this list focuses on markdown, although some of these tools, such as Vale~\cite{vale}, include support for a range of other formats (\eg{} reStructuredText). 
    See supplement for additional details. 
    }
    \Description{
A comparison table shows different linting tools and the quality checks  they support compared to LintMe tool. In total there are 13 rows for 13 different linting tool: markdownlint, mdformat, remark-lint, textlint, proselint, vale, content-linter, alex, write-good, standard-readme, awesome-lint, and grammarly. 
The upper part of the table shows rules listed as vertical labels along columns, with subgroup headings placed below them. This include, accuracy and reliability, consisting of rules such as execution, code formatting, link formatting, and link validation. Consistency consists of rules such as formatting/styling, structure, and terminology. Security consists of rules such as sensitive content and best practices. 
Timeliness stands alone without group. 
Conciseness consists of rules such as section completeness, repetition, sentence length, and too concise.
Accessibility consists of rules such as navigation, no-jargon, visual clarity, alt text, and cross platform. 
Availability and objectivity also stand alone without groups. 
Reputation consists of rules on hate words. 
Quality checks consists of rules such as spelling and grammar rules. 
Appearance consists of rule on content variety. 
Other rules includes date related, end with new line, emoji lint, and translations.
In the table, if a linting tool can flag rule, it is marked as Y; partial support is marked as P; and not supported marked as N. Colors reinforce this: Y with green, P with yellow, and N with red. From this table, if the linting tools is able to flag one of the rules, it will have Y, partially support P and not supported as N. With colors also telling the same thing with Y green, P yellow, N red. Most tools cover only handful of rules, by contrast, LintMe is the only tool that supports every rule with Y across the columns.
    }
        \label{tab:design-space}
\end{figure*}

%% file: figures/operators-table.tex
\begin{table*}[t]
\centering
\caption{A sample of the 21 operators available in \system{}. A full list of rules and additional details are available in the system documentation at \asLink{https://lintme.netlify.app/readme.md}{lintme.netlify.app/readme.md}. 
}
\label{tab:operators}
\Description{A comparison table shows different linting tools and the quality checks  they support compared to LintMe tool. In total there are 13 rows for 13 different linting tool: markdownlint, mdformat, remark-lint, textlint, proselint, vale, content-linter, alex, write-good, standard-readme, awesome-lint, and grammarly. 

While each column represent a list of rules with their subgroups. This include, accuracy and reliability, consisting of rules such as execution, code formatting, link formatting, and link validation. Consistency, consisting of rules such as formatting/styling, structure, and terminology. Security, consisting of rules such as sensitive content and best practices. Timeliness, which stands alone without group. Conciseness, consisting of rules such as section completeness, repetition, sentence length, and too concise.
Accessibility, consisting of rules such as navigation, no-jargon, visual clarity, alt text, and cross platform. Availability and objectivity, also stands alone without group. Reputation, consist rules on hate words. Quality checks, consisting of rules such as spelling and grammar rules. Appealness consist of rule on content variety. Other rules includes, date related, end with new line, emoji lint, and translations.

In the table, if a linting tool can flag rule, it is marked as Y; partial support is marked as P; and not supported marked as N. Colore reinforce this: Y with green, P with yellow, and N with red. From this table, if the linting tools is able to flag one of the rules, it will have Y, partially support P and not supported as N. With colors also telling the same thing with Y green, P yellow, N red. Most tools cover only handful of rules, by contrast, LintMe is the only tool that supports every rule with Y across the columns.
}
\setlength{\tabcolsep}{8pt}
\renewcommand{\arraystretch}{1.2}
\footnotesize
\begin{tabular}{|p{7.0cm}|p{8cm}|}

\hline
\textbf{Name} & \textbf{Description}  \\
\hline
\operator{compare}(baseline, against, comparison\_mode, similarity\_method, threshold) &
Compares outputs from two earlier steps either structurally (missing/extra items) or by similarity scoring. 
 \\
\hline
\operator{count}() &
Aggregates counts from the previous step’s extracted results and summarizes them by scope (\eg{} per line, per paragraph, or per collection). 
\\
\hline
\operator{customCode(code)} &
Executes user-provided JavaScript. Useful for bespoke checks or quick experiments that do not fit a built-in operator. 
 \\
\hline
\operator{evaluateUsingLLM}(model, ruleDefinition) &
Evaluates the current document against a provided rule definition using an LLM and returns PASS or FAIL with diagnostics. 
 \\
\hline
\operator{extract}(target, scopes) &
Finds Markdown nodes or text matches (including built-ins like emoji/newline/date and regex patterns) and returns them grouped by requested scopes. 
 \\
\hline
\operator{fetchFromGithub}(repo, branch, fileName, fetchType, metaPath, useCustomMetaPath) &
Fetches README files (paths or content) or repository metadata from a GitHub repo via a backend endpoint. 
 \\
\hline
\operator{regexMatch}(patterns, mode, scope) &
Flags lines that DO match (mode: ``match'') or do NOT match (mode: ``unmatch'') one or more regular expressions. Can run on the whole document or on the previous step's structured output. 
\\
\hline
\operator{search}(query, scope) &
Finds lines or values containing one or more comma-separated terms. Can scan the whole Markdown document or walk the previous step’s structured output. 
\\
\hline
\operator{threshold}(conditions) &
Compares computed metrics from a previous step (\eg{} count or length) against threshold rules per scope and reports violations. \\
\hline
\operator{execute}(timeout) &
Runs inline commands found in the README. Prefers the previous extract (target: inlineCode) output when available; otherwise scans backtick code spans. \\
\hline
\end{tabular}
\end{table*}

%% file: appendix.tex
\onecolumn

\section{Appendix}

In this appendix, we include a variety of materials that provide additional content or context to this work, but were not necessary to be included directly in the paper. 
Below we include our interview instrument (\secref{sec:interview-guide}), details about our LLM comparison study (\secref{sec:llm-eval-details}), and prompts we used throughout the system (\secref{sec:prompts}).
\autoref{tab:doc-rules} provides a listing of the lint rules used in the LLM replacability study and the user study.
In addition to these materials, we include the spreadsheets used to analyze the linters (giving Fig. 1), to analyze the results of our LLM comparison study (which yielded Fig. 5), and our recipe case study (yielding Fig. 6) are available at \osf{}. As noted elsewhere, our system can be found at \asLink{https://lintme.netlify.app/}{lintme.netlify.app} and the code for it can be found at \github{}.






\section{Interview Instrument}
\label{sec:interview-guide}

\subsection*{Purpose} Elicit practices and expectations for \readme{} documentation and evaluate the usefulness and workflow fit of \system{}.
\subsection*{Session Logistics}
\begin{itemize}
  \item Duration: 50--60 minutes (recorded with permission).
  \item Mode: Remote (screen shared).
  \item Materials: \asLink{https://lintme.netlify.app/}{lintme.netlify.app}; participant's \readme{} (optional).
  \item Compensation: Digital gift card sent to participant's provided email.
\end{itemize}

\subsection*{Opening Script (1--5 min)}  
Introductions, a brief overview of the tool, and an outline of the session agenda.  

\subsection*{Warm-Up (3--5 min)}
\begin{enumerate}
  \item Briefly describe your background with documentation and typical projects (coursework, machine learning, corporate, software development).
  \item What do you expect in a \emph{good} \readme{}? What makes one \emph{bad}? 
\end{enumerate}

\subsection*{Guided Demo (10--15 min)}
\begin{enumerate}
  \item Walk through the main features: loading Markdown files, loading rules, editing rules, running rules or presets, and viewing results.  
  \item Demonstrate how to create custom rules (e.g., using Generate Rule with LLM) and apply quick fixes.  
\end{enumerate}

\subsection*{Hands-On Tasks (10--20 min)}
Ask the participant to share their screen and think aloud while working.  
They may use their own \readme{} file or select a sample provided in the tool.  
Tasks include:  
\begin{itemize}
  \item running a single rule or a preset to identify and fix issues of their choice, and  
  \item using the "Generate Rule" feature to create and apply a custom rule.  
\end{itemize}

\subsection*{Core Questions (10--20 min)}
These are the rules we use to structure our questions. We adjust them and add follow-up questions depending on the users' responses.
\paragraph{First Impressions \& Learnability}
\begin{enumerate}
  \item How did you feel about the tool and the process of creating or modifying a rule?
\end{enumerate}

\paragraph{Usefulness \& Workflow Fit}
\begin{enumerate}\setcounter{enumi}{1}
  \item Would you integrate \system{} into your documentation workflow?
  \item Would you recommend \system{} to someone else?
  \item What other lint rules are you interested in?
\end{enumerate}

\paragraph{Community Standards \& Iteration}
\begin{enumerate}\setcounter{enumi}{4}
  \item What do you consider to be important elements of a high-quality \readme{} in your field?
  \item Do you think everyone in your field agrees on the elements that make up a high-quality \readme{}? Why or why not?
\end{enumerate}

\subsection*{Closing (1--2 min)}
Any final thoughts on something we did not cover?

\section{LLM Evaluation Details}
\label{sec:llm-eval-details}

In our replaceability evaluation, we conducted a comparison with a relatively naive LLM Comparison. Below, we include the prompts for the modes considered. At \osf{} we include the results tables of this evaluation.

\subsection{\freePrompt{}}

\textbf{System Prompt}: \texttt{You are an experienced software developer and technical writer.\\
Your job is to review a \readme{} file and suggest improvements.}\newline
\noindent{}\textbf{User Prompt}: \texttt{
Here is a \readme{}. Please:\\
1) Apply your own judgment and identify any improvements that could be made to improve the \readmes{} quality.\\
2) Output only a list in the format:\\
   -Description of the needed change
3) Do not rewrite or modify the \readme{} itself.\\
--- \readme{} START ---\\
\readme{} file\\
--- \readme{} END ---
}

\subsection{\rulesProvided{}}

\textbf{System Prompt}: \texttt{You are an experienced software developer and technical writer. \\ 
Your job is to review a \readme{} file and suggest improvements}\\\\
\noindent{}\textbf{User Prompt}: \texttt{
Here is a \readme{}. Please:\\
1) Review each of the rules listed below and, based on them, identify any improvements that could be made to improve the \readmes{} quality.
Do not invent new rules. Only apply the ones provided.\\
Rules:\\
{rules text}\\
2) Output only a list in the format:\\
   Rule X: [description of the needed change]\\
3) Do not rewrite or modify the \readme{} itself.\\
   Do not output anything for rules that do not require changes.\\
--- \readme{} START ---\\
\readme{} file\\
--- \readme{} END ---
}

\section{Prompts From The Tool}
\label{sec:prompts}
The tool invokes large language models (LLMs) at several stages. Each LLM operator behaves like any other operator in the pipeline and is defined in YAML. When invoked, the user-provided prompt is combined with a predefined system prompt and sent to the LLM. This section documents the key prompts used by the tool.

\subsection{rule-generator}
\label{sec:rule-generator}
The generate rule feature is used to compose linting rules by combining a system prompt with the \texttt{Operators Catalog}. The catalog contains JSON schemas describing each operator, including the fields \texttt{id}, \texttt{description}, \texttt{allowedFields}, \texttt{fields}, and \texttt{examples}. These schemas constrain how rules can be generated. For illustration, the example below includes only a single operator, \texttt{isLinkAlive}, rather than the full catalog.
\newline

\noindent{}\textbf{System Prompt}: \texttt{You are a \readme{} linter rule designer. Return YAML ONLY with keys: rule, description, pipeline. Use ONLY operator ids listed in the OPERATORS CATALOG. Use only the fields listed for each operator under Allowed fields. Do not add fields that are not listed. Use enum values exactly as listed when present. Include a field only if it appears in Allowed fields for that operator. YAML only. No prose. No code fences. Prefer built-in operators; Do not use customCode.
}
\\\\
\noindent{}\textbf{Operators Catelog:}

\noindent{}\textbf{id:} \texttt{isLinkAlive} \\
\noindent{}\textbf{description:} \texttt{Checks all external links found in the Markdown content to verify they are reachable and return an allowed HTTP status code.} \\
\noindent{}\textbf{allowedFields:} \texttt{[timeout, allowed\_status\_codes]} \\
\noindent{}\textbf{fields:}
\begin{itemize}
  \item \textbf{name:} \texttt{timeout} \\
  \textbf{type:} \texttt{integer} \\
  \textbf{default:} \texttt{5000} \\
  \textbf{description:} \texttt{Request timeout in milliseconds for each link check.}
  \item \textbf{name:} \texttt{allowed\_status\_codes} \\
  \textbf{type:} \texttt{array} \\
  \textbf{default:} \texttt{[200, 204, 301, 302, 307, 308]} \\
  \textbf{description:} \texttt{List of acceptable HTTP status codes for a link to be considered alive.}
\end{itemize}

\noindent{}\textbf{examples:}
\begin{verbatim}
operator: isLinkAlive
timeout: 3000
allowed_status_codes:
  - 200
  - 301
  - 302
\end{verbatim}

\noindent{}\textbf{User Prompt}: \texttt{ Idea: each markdown must have 2 headings.
Return the final YAML now.
}

\subsection{fixUsingLLM operator prompt}
\label{sec:operator-prompt}

The \texttt{fixUsingLLM} operator uses an LLM to correct Markdown content that violates a specific rule automatically. The user prompt specifies what fix to apply, while the system prompt enforces strict constraints to avoid unintended modifications.\\
\newline
\textbf{System Prompt}: \texttt{
You are a Markdown linter. Your job is to fix ONLY the issues that violate the specific rule defined below.\\
Rule Definition (YAML):
\{ruleYaml\}\\
\\
Operator-Specific User Prompt:
\{prompt\}\\
\\\
Diagnostics from Previous Steps:
\{diagnosticText\}\\
\\
Markdown Document:
\{ctx.markdown\}\\
\\
Very Important Instructions:\\
- ONLY fix issues that are directly and clearly covered by the rule above.\\
- DO NOT make any changes based on grammar, tone, inclusivity, or clarity unless the rule explicitly calls for it.\\
- DO NOT invent improvements or infer intent not stated in the rule.\\
- If the Markdown content does NOT violate the rule, return the original input exactly as is — unchanged.\\
- Behave like a deterministic function: same input → same output.\\
- If there is even slight ambiguity in whether something violates the rule, you must not change it.\\
- DO NOT change headings, formatting, phrasing, or terms unless they clearly break the rule.\\
- Preserve exact trailing newlines if present in the original Markdown.\\
\\
Output Format:\\
- ONLY include the corrected (or unmodified) Markdown \textbf{below} the marker.\\
- NEVER include explanations, comments, or wrap it in a code block.\\
\\
---FIXED MARKDOWN BELOW---
}

\subsection{evaluateUsingLLM operator prompt}
The \texttt{evaluateUsingLLM} operator asks an LLM to evaluate whether a given Markdown document violates a specified rule. The output must strictly conform to a pass/fail format, including line references and suggested fixes when applicable.
\\\\
\textbf{System Prompt}: \texttt{
You are a Markdown rule checker. Your job is to determine if the given Markdown violates the provided rule.\\
Rule Definition (YAML):
\{ruleYaml\}\\
\\
Intermediate Outputs:
\{operatorOutputs\}\\
\\
Previous Diagnostics:
\{diagnosticText\}\\
\\
Markdown Document:
\{ctx.markdown\}\\
\\
Instructions:\\
1. First, determine if the Markdown violates the rule.\\
2. If it **does not**, reply exactly as follows:\\
\\
**PASS**\\
3. If it **does**, reply exactly in this format:\\
\\
**FAIL**\\
Line(s): [list affected line numbers]\\
Issue: [brief summary of the issue]\\
Suggestion: [suggest a fix using natural language]\\
\\
Respond only in the above format — no code blocks, no additional comments.\\
}
\input{figures/evaluated-rules.tex}

%% file: figures/evaluated-rules.tex
\clearpage
\begin{table}[p]
\small
\caption{The set of rules used during the LLM Evaluation of README files.}
\label{tab:doc-rules}
\centering
\begin{tabular}{|p{0.41\linewidth}|p{0.55\linewidth}|}
\hline
\textbf{Rule} & \textbf{Description} \\
\hline
\lintwithoutul{ambiguity-understandability-check(llm)} & Require clear, unambiguous, and audience-appropriate instructions. \\
\hline
\lintwithoutul{best-practices-guidelines(llm)} & Enforce security and best practice guidelines. \\
\hline
\lintwithoutul{check-back-to-top-link-presence} & Require a “Back to Top” link for navigation. \\
\hline
\lintwithoutul{citation-bibtex-present} & Require BibTeX citations where applicable. \\
\hline
\lintwithoutul{code-block-consistency} & Enforce consistent formatting of code blocks. \\
\hline
\lintwithoutul{code-block-consistency-bash} & Enforce consistent formatting for bash blocks. \\
\hline
\lintwithoutul{code-block-execution(llm)} & Require code blocks to be valid, complete, and executable. \\
\hline
\lintwithoutul{code-block-language-check} & Require language tags on all code blocks. \\
\hline
\lintwithoutul{compare-markdown-renderings} & Require consistent rendering across Markdown engines. \\
\hline
\lintwithoutul{compare-readme-similarity} & Require $\geq$80\% similarity with previous README versions. \\
\hline
\lintwithoutul{consistent-external-link-format} & Require consistent formatting of external links. \\
\hline
\lintwithoutul{consistent-heading-format} & Require consistent heading formatting. \\
\hline
\lintwithoutul{consistent-list-format} & Require consistent list formatting. \\
\hline
\lintwithoutul{custom-code-hate-speech} & Disallow hate or discriminatory language. \\
\hline
\lintwithoutul{custom-code-sentence-length} & Limit sentence length in custom code. \\
\hline
\lintwithoutul{date-validation-lint} & Enforce correct and consistent date formats. \\
\hline
\lintwithoutul{demo-link-required-lint} & Require a demo link. \\
\hline
\lintwithoutul{demo-link-required-lint-interactive-systems} & Require a demo link for interactive systems. \\
\hline
\lintwithoutul{detect-duplicate-sentences} & Disallow duplicate sentences. \\
\hline
\lintwithoutul{detect-hate-speech-lint(llm)} & Disallow biased or profane language; suggest respectful alternatives. \\
\hline
\lintwithoutul{detect-sensitive-secrets} & Disallow secrets, keys, or tokens. \\
\hline
\lintwithoutul{disallow-consecutive-duplicate-words} & Disallow consecutive duplicate words. \\
\hline
\lintwithoutul{enforce-emoji-limit} & Limit emoji usage at document, paragraph, and line levels. \\
\hline
\lintwithoutul{enforce-newline-at-eof} & Require a newline at end of file. \\
\hline
\lintwithoutul{ensure-neutral-tone(llm)} & Require a neutral, credible, and professional tone. \\
\hline
\lintwithoutul{ensure-readme-is-present} & Require a README in standard location. \\
\hline
\lintwithoutul{fix-spelling-and-grammar(llm)} & Enforce correct spelling and grammar using an LLM. \\
\hline
\lintwithoutul{github-contributor-check} & Require contributor or maintainer details. \\
\hline
\lintwithoutul{heading-list-formatting-checks} & Require proper heading hierarchy and list formatting. \\
\hline
\lintwithoutul{jargon-explanation-check(llm)} & Require explanations for jargon. \\
\hline
\lintwithoutul{limit-heading-count} & Limit heading depth. \\
\hline
\lintwithoutul{limit-word-repetition} & Limit excessive word repetition. \\
\hline
\lintwithoutul{minimum-readme-length} & Require sufficient README length. \\
\hline
\lintwithoutul{translations} & Require translations for non-English text. \\
\hline
\lintwithoutul{no-unreachable-links} & Disallow broken links. \\
\hline
\lintwithoutul{objectivity-check(llm)} & Require objective and unbiased content. \\
\hline
\lintwithoutul{quick-start-section-check} & Require a quick start section in long docs. \\
\hline
\lintwithoutul{require-alt-text-for-images} & Require alt text for images. \\
\hline
\lintwithoutul{rich-content-check} & Require varied, structured content (headings, images, emphasis). \\
\hline
\lintwithoutul{section-completeness(llm)} & Require all standard documentation sections. \\
\hline
\lintwithoutul{section-completeness(llm)-interactive-systems} & Require full sections for interactive systems. \\
\hline
\lintwithoutul{sentence-length-limit} & Limit sentence length. \\
\hline
\lintwithoutul{sentence-length-limit-datasets} & Limit sentence length in dataset docs. \\
\hline
\lintwithoutul{table-of-contents-check} & Require a table of contents with links. \\
\hline
\lintwithoutul{terminology-consistency(llm)} & Enforce consistent terminology. \\
\hline
\lintwithoutul{text-contrast-compliance} & Require sufficient text contrast. \\
\hline
\lintwithoutul{timeliness-check-version-update(llm)} & Require up-to-date version information. \\
\hline
\lintwithoutul{validate-inline-commands} & Require valid inline command formatting. \\
\hline
\lintwithoutul{validate-internal-links-to-headings} & Require valid internal heading links. \\
\hline
\lintwithoutul{validate-link-formatting} & Require proper link formatting. \\
\hline
\lintwithoutul{value-added-lint} & Require added value (guides, resources). \\
\hline
\end{tabular}
\end{table}
\clearpage